\documentclass{jfm}

\usepackage{graphicx}
\usepackage{newtxtext}
\usepackage{newtxmath}
\usepackage{natbib}
\usepackage{hyperref}
\hypersetup{
    colorlinks = true,
    urlcolor   = blue,
    citecolor  = black,
}

\newcommand{\RomanNumeralCaps}[1]
\linenumbers

\usepackage{subcaption}
\usepackage{placeins}
\usepackage{multirow}

\usepackage{color}
\usepackage{xcolor}

\title{Perpendicular rod-airfoil aeroacoustics: experiments and modelling of interaction noise}

\author{Marios I. Spiropoulos\aff{1}
  \corresp{\email{marios-ioannis.spiropoulos@ensma.fr}},
 Filipe R. Amaral\aff{1,2}, Florent Margnat\aff{1}, Vincent Valeau\aff{1}, \and Peter Jordan\aff{1}}

\affiliation{\aff{1}DFTC-2AT, Institut Pprime, CNRS – Université de Poitiers  – ISAE-ENSMA, UPR 3346, 11 Boulevard Marie et Pierre Curie, 86073 Poitiers, Vienne, France.\aff{2}Divisão de Aeronáutica, Instituto Tecnológico de  Aeronáutica, Praça Marechal Eduardo Gomes, 50, 12228-900 São José dos Campos, São Paulo, Brazil.}

\begin{document}

\maketitle

\begin{abstract}
During the phase of landing, an important aircraft-noise source emanates from the interaction of the landing-gear-wake with the deployed flap. In the present work we cast this problem in an academic framework, by studying a simplified configuration that consists of a cylinder placed upstream and perpendicularly to a symmetrical NACA-0012 airfoil. An experimental campaign is conducted, followed by modelling approaches to explore the flow phenomena associated with the acoustic field. Simultaneous acoustic and stereoscopic Time-Resolved Particle Image Velocimetry measurements are taken, to study the sound and flow-fields generated by the interaction of the cylinder-wake with the downstream airfoil, when the spans of the two objects are orthogonally aligned. The experimental data highlight the three-dimensional nature of the problem. The maximum sound pressure levels are obtained at frequencies close to $\mathrm{St}\equiv f d/U \approx 0.38$ (cylinder's drag fluctuation frequency), where also the maximum linear coherence between the acoustic and cylinder-span-oriented fluctuation velocity is observed, demonstrating that the measured acoustic-field is an outcome of the three-dimensional cylinder-wake. Powell-Howe vortex-sound theory combined with an acoustically-compact Green function for the NACA-0012 are employed for the aeroacoustic modelling. A linearised source-term based on the analysed experimental data is used as input to estimate the acoustic field and identify the acoustically important coherent structures of the flow-field. A reasonable agreement is obtained between the sound field estimations and the measurements. To further explore the mechanisms of sound generation, a semi-empirical source-model, informed by the experimental data, is proposed, based on Fourier-modes in the cylinder's span direction.
\end{abstract}

\section{Introduction}
\label{sec:Intro}

Aircraft noise is divided mainly in two categories; i) the noise generated by the jet-engines and ii) the interaction of the flow with the aircraft's structure (airframe noise) which becomes significant during the phase of landing when the high lift devices are deployed. The interaction of the landing-gear wake with the deployed flap is considered a major airframe noise component, which can exceed the landing gear-noise itself \citep{oerlemans2004experimental}. Experiments were conducted to investigate the far-field acoustic radiation \citep{oerlemans2004experimental, pott2017study}, while flow measurements and numerical simulations  \citep{khorrami2015comparative} were carried out to study the aerodynamic field in the vicinity of the landing gear and the downstream flap. However, the mechanism of sound generation due to the interaction of the landing-gear-wake with the surface of the flap is poorly understood. In the present work, a simplified configuration, which consists of a circular cylinder and a downstream airfoil, with their spans orthogonally aligned, is proposed to mimic the landing-gear/flap system and allow the investigation of the key mechanisms related to sound generation via detailed experiments and semi-empirical models.

The latter can be classified in the wider category of  noise generated due to turbulence-airfoil interactions which is found in many industrial applications, to name a few examples: wind-turbine generated noise when the turbine blades interact with atmospheric turbulence \citep{LIU2017311}, ingestion noise of turbulence impinging on rotating blades systems \citep{Raposo_Azarpeyvand_2024}, aircraft and helicopter noise \citep{schlinker1983rotor}. 

In its simplest approximation, the present problem can be reduced into the aeroacoustics of a single line-vortex orthogonally impinging on a downstream airfoil which has been treated theoretically \citep{amiet1986airfoil,howe1988contributions, howe1989unsteady} and experimentally \citep{schlinker1983rotor,ahmadi1986experimental}. In particular, \citet{schlinker1983rotor} investigated the acoustic field of a vortex, cut at different angles by a large rotor-blade, and showed that almost no sound is generated when the vortex impinges orthogonally on the airfoil. \citet{ahmadi1986experimental} reproduced the previous experiments and demonstrated that when the unsteady velocity parallel to the vortex-axis is small enough, the acoustic spectra drop significantly with respect to cases where the vortex is cut at an arbitrary angle by the blade. \citet{howe1988contributions, howe1989unsteady}, using vortex-sound theory demonstrated theoretically the previous experimental observation by concluding that when a vortex is cut orthogonally by a downstream airfoil, no sound is generated unless there is an unsteady velocity parallel to the vortex axis.

The modelling of such problems relies mainly on Amiet's formulation and its extensions \citep{amiet1975acoustic,roger2005back} or similar formulations \citep{Howe2001,Hales_2023} where the acoustic field is given provided that the turbulent spectra are used as input. \citet{Roger_Extensions} discussed the use of those models in applications where the linearised unsteady aerodynamic theory is valid (small camber and thickness of airfoil), while also extensions are provided for rotating blades of finite span and the effect of the sweep angle \citep{roger2014vortex, quaglia20173d}. For instance, \citet{Wang_Wang_Wang_2021} computed the acoustic field when the wake of a circular cylinder is ingested by a motor. The turbulent statistics around the blades were informed by compressible large eddy simulations (LES) and the sound field was propagated by using Amiet's formulation and Sears theory to account for the transfer function between the unsteady lift and the generated upwash velocity. 

 A benchmark case of wake-airfoil-interaction problems is the classical rod-airfoil experiment of \citet{jacob2005rod}, which was further explored using numerical computations of the flow and the acoustic far fields \citep{berland2010numerical, Jiang_Mao_Deng_Liu_2015}. \citet{jacob2005rod} have shown that sound waves are produced due to the interaction of the vortex-street, shed in the cylinder's wake, with the leading edge of the airfoil, while \cite{casalino2003prediction} highlighted the importance of the airfoil-span extent of the unsteady aerodynamic field near the airfoil. The latter problem contains the same elements as the present study (cylinder's wake interacting with a downstream airfoil), however, it is shown in a recent study that for the perpendicular rod-airfoil configuration, the mechanisms of sound generation substantially differ due to the different orientation of the cylinder and the airfoil which leads to the orthogonal cut of the shed von Kármán vortices by the downstream airfoil \citep{do_amaral_perpendicular_2025}.

Based on the existing literature, in the purely two dimensional case of a cylinder's wake interacting orthogonally with the downstream airfoil, it would be expected that the measured acoustic field would be negligible. However, as will be discussed in what follows, the flow-field is three dimensional and thus the generated sound field exceeds the one of the airfoil or the cylinder alone. Therefore, the present work can be regarded as a follow-up study of the work of \cite{do_amaral_perpendicular_2025}, where the acoustically important flow structures where extracted via post-processing techniques performed on a large experimental data-base. In the present work we use the same data-base aiming to: i) document and report the main experimental observations in terms of the flow and acoustic fields; ii) propose a simplified, semi-empirical model based on vortex-sound theory to further explore the physics of sound generation and iii) propose a hypothesis of the origin of the flow structures associated with the sound field.

The strategy followed to model the acoustic field and explore the physics of the interaction noise diverges from the classical approach of considering the turbulent spectra in the vicinity of the airfoil as the sound-source. We are rather interested in identifying the acoustically important coherent structures of the turbulent flow field that may be linked with other physical flow processes such as instabilities in the cylinder's wake. Aeroacoustic models based on coherent structures have been successfully applied in jet-noise problems \citep{jordan2013wave} and more recently in bluff-body noise problems \citep{sano2019trailing, Abreu_Tanarro_Cavalieri_Schlatter_Vinuesa_Hanifi_Henningson_2021, Simon_wavepackets, prinja2025experimentally,Zhenyang_wavepackets}. Motivated by the simplicity of the wave-packet models proposed to understand jet-noise \citep{ CAVALIERI20146516, cavalieri2019wave}, we propose a similar semi-empirical approach to represent the aerodynamic source as wave-structures, decomposed in cylinder-span aligned waves of different wavelengths. The latter are used to investigate the effect of the topological properties of the flow field to the sound-generation process.

The remainder of the paper is organised as follows: In section \ref{sec:Experiments} we describe the experimental campaign carried out to characterise the flow and acoustic fields of the vertical cylinder's wake-airfoil interaction noise by performing synchronous flow and acoustic measurements. In the second part of the paper (section \ref{sec:TVS}), we combine the experimental results with Powell-Howe's vortex-sound theory to derive a simplified, linearised experimentally informed model. In section \ref{sec:semi_empirical_models} we combine this model with a data-based, semi-empirical source model  to investigate the acoustically important flow structures and their corresponding length scales.  Finally, an assumption regarding a possible origin of the source-term is discussed opening the question, whether the observed acoustically important flow structures are associated with a secondary instability of the cylinder's wake.

\section{Experimental investigation}
\label{sec:Experiments}
\begin{figure}
\centering
\includegraphics[width=0.8\textwidth]{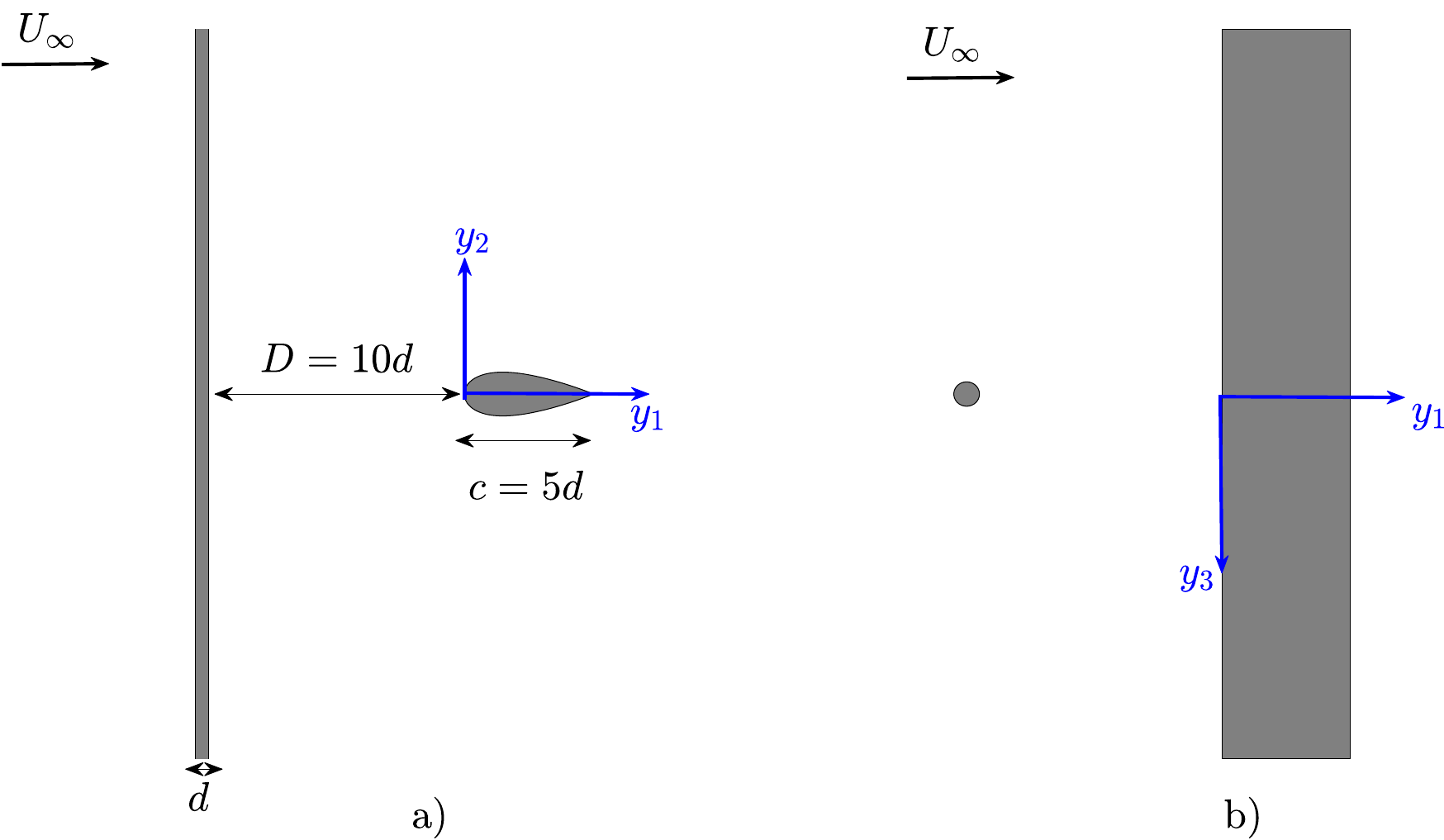}
\caption{Coordinate system of the experimental set-up, side (a) and top (b) view. The distance between the cylinder and the airfoil is $D=10d$, while the distance between the nozzle-exit and the cylinder is $35.5d$.}
\label{setup}
\end{figure}

The first part of our study deals with the experimental investigation conducted in Pprime Institute's anechoic wind tunnel \textit{Bruit Environnement Transport Ingénierie} (BETI). BETI is an Eiffel-type wind tunnel with an open test section of $0.7 \times 0.7$ m$^2$, turbulence intensity $<1\%$ and reaches a maximum flow-speed of $60$ m/s. The volume of the anechoic chamber and its cut-off frequency are $90$ m$^3$ and $200$ Hz, respectively. The experimental set-up used in the present work is the same of the one presented by \cite{do_amaral_perpendicular_2025} with some slight modifications. For a complete view of the experimental campaign the reader may find complementary details in \citep{do_amaral_perpendicular_2025}. The set-up consists of a cylinder with diameter ($d=0.02$ m) placed at a distance $l=35.5d$ from the nozzle exit and a NACA-0012 airfoil with chord of $0.1$ m $\left(c=5d \right)$ located at $D=10d$ further downstream from the cylinder, as shown in figure \ref{setup}. The coordinate system is defined as well in figure \ref{setup}, where $y_1, y_2, y_3$ denote the streamwise, cylinder-span-aligned, airfoil-span aligned directions respectively. The flow speed is fixed at $U_\infty=30 $ m$/$s, corresponding to a Mach number $\mathrm{M}=0.09$. For the airfoil, this velocity leads to $\mathrm{Re} \approx 2 \times 10^5$ in air at $20 \mathrm{^\circ C}$. At this value, tonal trailing-edge noise is expected \citep{paterson1973vortex}, which is suppressed here by tripping the  boundary-layer transition to turbulence. The tripping device consists of a sandpaper band, placed at $5 \%$ of the chord from the leading edge, on the suction and pressure side. For the cylinder, the Reynolds number is $\mathrm{Re}  \approx 4 \times 10^4$, corresponding to the subcritical turbulent regime. At this value, the spanwise coherence length is about five diameters at the lift fluctuation frequency, with no significant dependency upon the specific value of $\mathrm{Re}$ \citep{margnat2023cylinder}. Consequently, this setting corresponds without loss of generality to a qualitative analysis of a cylinder-wake impinging on the airfoil. Detailed, synchronised measurements took place to characterise the flow and acoustic fields and identify links between the velocity components and the measured sound field. In what follows we provide a discussion regarding the experimental campaign, complementary to the one presented in the recent study of  \cite{do_amaral_perpendicular_2025}.

\subsection{Flow and acoustic measurements}
\label{subSec:Measurements}
The flow and acoustic fields are recorded using stereoscopic Time-Resolved Particle Image Velocimetry measurements (sTR-PIV) and a set of analog microphones located around the cylinder-airfoil configuration respectively.

For future reference we fix our notation as follows. Let $q$ be an arbitrary quantity: i) all unsteady fluctuation quantities in the time domain, will be denoted as $q'(t)$, ii) in the frequency domain as $\hat{q}(\mathrm{St})$ and iii) we will refer to time-averaged quantities as $\bar{q}$. The non-dimensional frequency (Strouhal number) used throughout the present work is defined as $$\mathrm{St}=\frac{fd}{U_\infty},$$ where $f$ the frequency in Hertz. Furthermore, the subscript "$i$" as in $q_i$ denotes the component of $q$ in the $i$-th direction, unless stated otherwise (for example when using the acoustic pressure $p_j$, in that case "$j$" corresponds to the number of the microphone). What is more, the source-related coordinates will be denoted by the vector $\vec{y}$, while the observer in the acoustic far-field will be described by the vector $\vec{x}$, both non-dimensionalised with the chord of the airfoil ($c$).

All quantities are used in their non-dimensional form unless stated otherwise. The velocity components ($\tilde u_i$) are non-dimensionalised as $$u_i= \frac{\tilde u_i}{U_\infty},$$ while the vorticity components as $$ \omega_i= \frac{\tilde \omega_i d}{U_\infty},$$ 

\noindent where $\tilde{\cdot}$ denotes dimensional quantities.
\subsubsection{Acoustic measurements}
\label{subsubSec:Acoustic_measurements}
\begin{figure}
\centering
\includegraphics[width=0.95\textwidth]{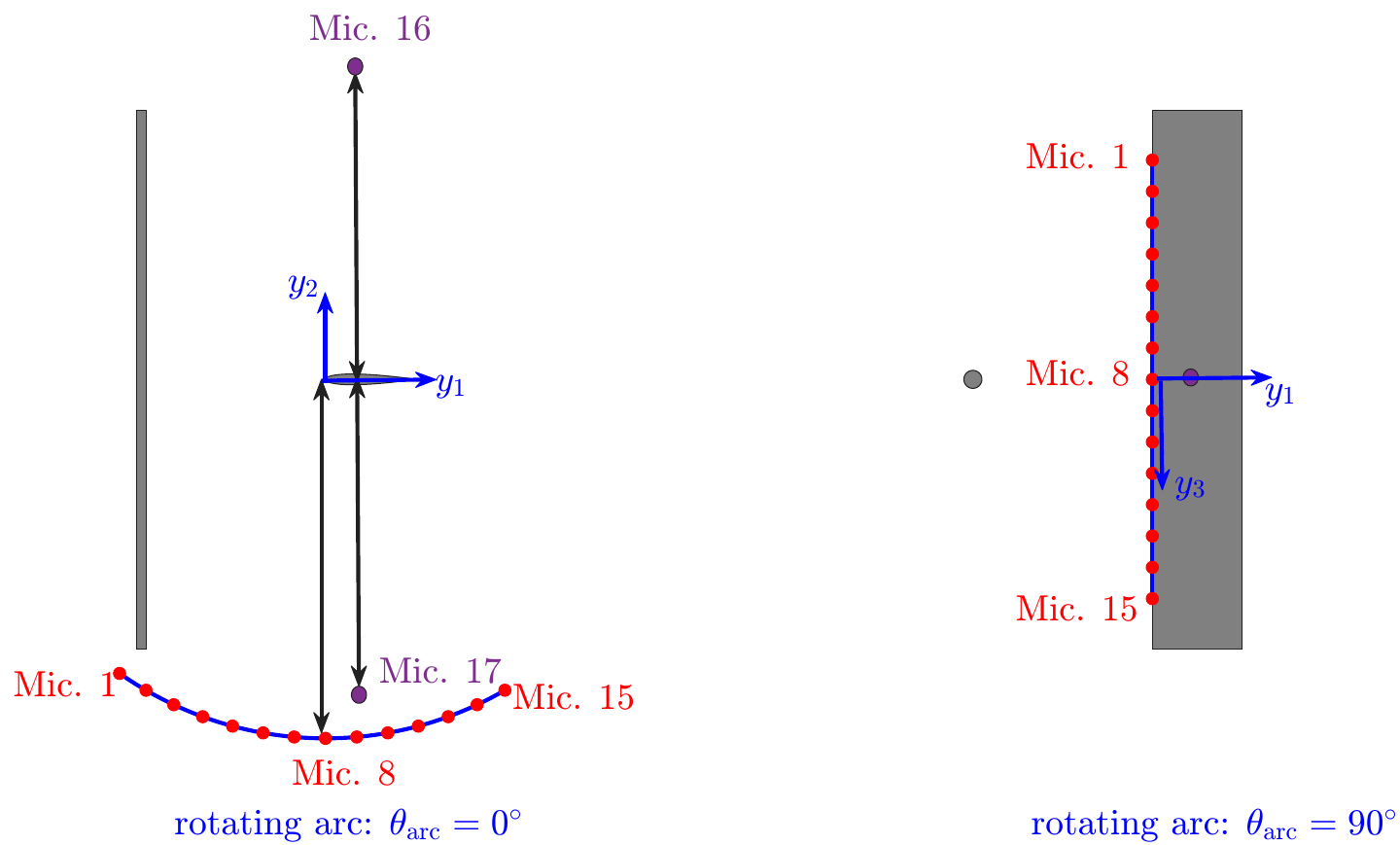}
\caption{Side (left) and top (right) view of the experimental set-up: The dots in magenda correspond to the analog microphones that were synchronised with the sTR-PIV-cameras. The red dots correspond to the microphones mounted on a rotating arc (blue arc) that was not synchronised with the PIV-cameras. At its initial position the arc of microphones is used to characterise the acoustic field below the airfoil in the $y_1-y_2$ plane (a) while a rotation of $90^\circ$ corresponds to the set-up shown in the $y_1-y_3$ plane (b). The positions of the microphones are not in scale.}
\label{Set_up_with_mics}
\end{figure}

The acoustic field has been characterised with analog RIOM ICP-microphones of $1/2$ inches placed above and below the airfoil as shown in figure \ref{Set_up_with_mics}. It is noted that microphones $16,17$ of figure \ref{Set_up_with_mics} were synchronised with the cameras used during the sTR-PIV-campaign to record simultaneously flow and acoustic measurements.

 To complement the measurements, a rotating arc of $15$ microphones corresponding to a circular sector of $70^\circ$ was placed below the leading edge of the airfoil. The arc was initially placed as shown in figure \ref{Set_up_with_mics} (left), to capture the directivity of the sound field below the airfoil. Then, the arc is rotated by $90^\circ$ so that the microphones are distributed along the span of the airfoil to investigate the sound radiation emanating from the cylinder's lift dipole [figure\ref{Set_up_with_mics} (right)]. The duration of the measurements was set at $T=15$s and the data were acquired at a sampling frequency of $f_s=50$ kHz, which corresponds to $75 \times 10^4$ samples.

\begin{figure}
\centering
\includegraphics[width=0.75\textwidth]{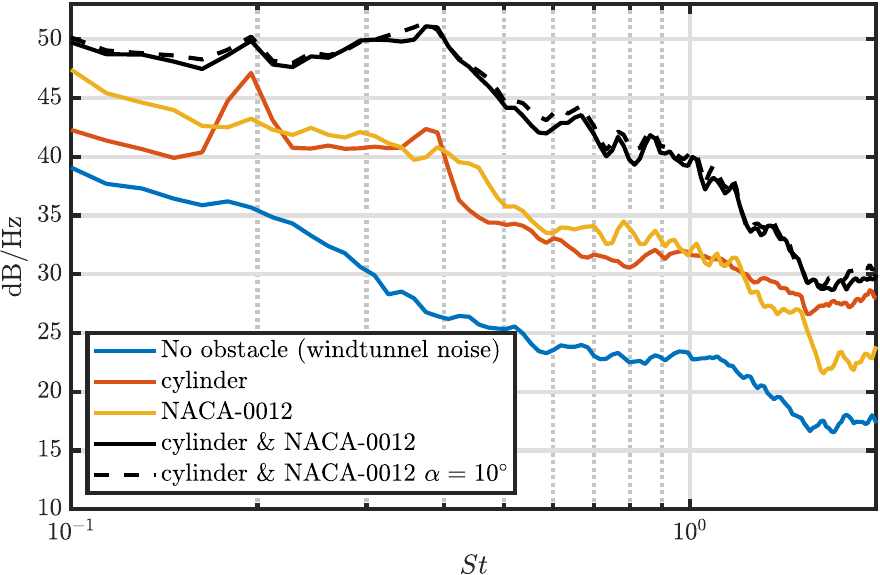}
\caption{Measured sound pressure levels (microphone 16 at $\vec{x}=(c/2,6.9c,0$) when no obstacle is placed in the wind tunnel (blue curve), the airfoil only (yellow), the cylinder only (orange) and both objects are placed in the wind tunnel (black lines). The solid and dashed black curves correspond to an angle of attack of the airfoil ($\alpha$) of $0$ and $10^\circ$ respectively. The flow-speed is set to $U=30$ m/s.}
\label{Interaction_verticalhorizontal}
\end{figure}

\begin{figure}
\centering
\includegraphics[width=0.6\textwidth]{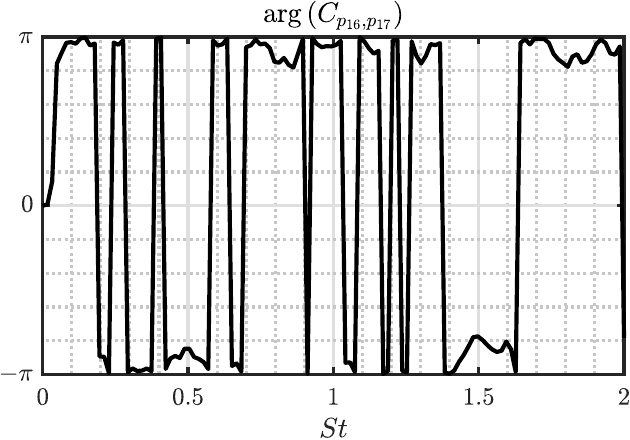}
\caption{Phase of the cross-spectral density of the microphones symmetrically placed at a distance above and below the airfoil at $\vec{x}_{\text{Mic.}_{16},\text{Mic.}_{17}}=(c/2, \pm 6.9c,0 )$.}
\label{Arg}
\end{figure}

The sound pressure level is defined as, $$\mathrm{SPL}=20 \log_{10} \left( \frac{\left< \hat{p}(f)\hat{p}^*(f) \right> }{p_\text{ref}^2} \right),$$
where $\left< \hat{p}(f)\hat{p}^*(f) \right> $ is the acoustic power spectral density and is evaluated via the Welch method using blocks of $N_\text{fft}=2048$ points and an overlap of $50 \%$ and $p_\text{ref}=20$ $\mu $Pa . The microphone located at $6.9c$ above the center of the airfoil (Mic. 16 on figure \ref{Set_up_with_mics}) was used to measure the sound field, when the cylinder only, the NACA-0012 only and both objects are placed in the wind tunnel and the acoustic spectra are shown on Figure \ref{Interaction_verticalhorizontal}. A $10$ dB increase is observed in the sound pressure levels generated by the interaction of the cylinder's wake with the downstream airfoil. Two peaks are observed, one that corresponds to the lift frequency of the cylinder flow $\left( \mathrm{St}=0.19 \right)$ and another one at the drag frequency $\left( \mathrm{St}=0.38 \right)$, which will be shown to be an important contributor to the interaction noise. Moreover, the angle of attack of the airfoil ($\alpha=10^\circ$) does not influence significantly the acoustic spectra and will therefore be excluded as a parameter of the problem. To verify the dipolar nature of the acoustic field we compute the phase of the cross-spectral density of the microphones located above and below the center of the airfoil (Mics. 16,17). Figure \ref{Arg} illustrates that in the frequency-band of interest there is a clear phase opposition indicating the existence of an acoustic dipole oriented in $y_2$.

\begin{figure}
\centering
\includegraphics[width=0.75\textwidth]{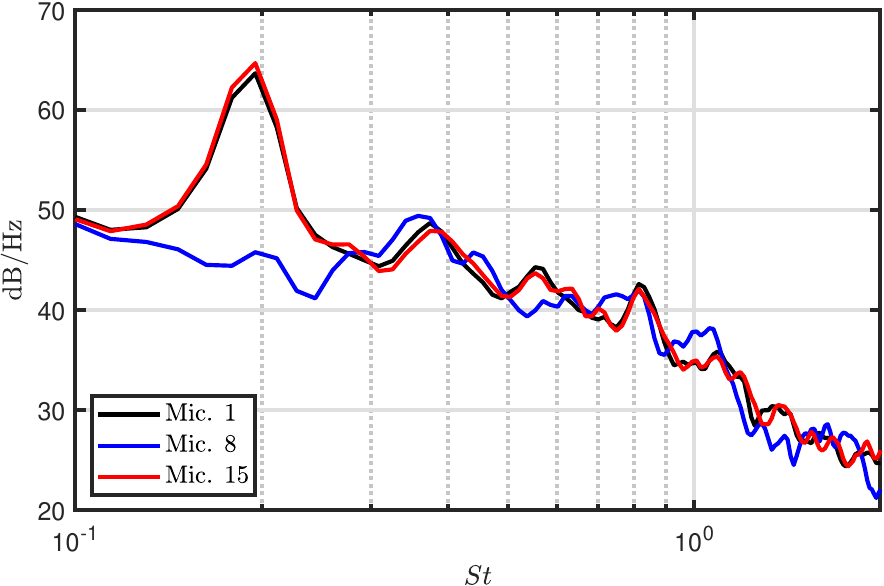}
\caption{Sound pressure levels obtained by the extreme microphones placed symmetrically along the span of the airfoil (Mics. 1,15) and the microphones located in the midspan, at a distance of $10c$ below the leading edge (Mic. 8). The microphones are placed according to figure \ref{Set_up_with_mics} b.}
\label{cylinder_noise}
\end{figure}

The current configuration differs from the classical rod-airfoil experiment, in the sense that the axis of the cylinder is placed orthogonally to the axis of the airfoil. \citet{jacob2005rod} have shown, for the case where the cylinder and airfoil spans are parallel, that the von Kármán vortex street interacting with the downstream airfoil leads to sound production and hence the main direction of sound generation is parallel to the unsteady lift acting on the cylinder and the airfoil. This is evident in the acoustic spectra that the authors presented, which contain the main frequency peaks related to the vortex-shedding frequency and its harmonics, spread over a wider range of frequencies. Due to the orientation of the current set-up, the three-dimensionality of the flow field is expected to give rise to two main noise sources,  one emanating from the cylinder in the $y_3-$ direction (parallel to the span of the airfoil) and a second, denoted as wake-airfoil interaction noise, oriented in the $y_2$-direction (parallel to the cylinder's span). These acoustic dipoles can be observed in the detailed characterisation of the acoustic field of \cite{do_amaral_perpendicular_2025}. As shown in figure \ref{cylinder_noise} the acoustic power spectra computed by the microphones above and below the mid-span (Mics. 1,15) of the airfoil (Fig \ref{Set_up_with_mics}b) have a broadband character with a strong peak at the vortex-shedding frequency ($\mathrm{St} = 0.19$) related to the cylinder's lift fluctuation frequency and a weaker one at $\mathrm{St} = 0.38$ related to the cylinder's drag fluctuation. Those unsteady forces on the cylinder result in the corresponding lift and drag dipoles that characterise cylinder noise \citep{inoue2002sound}. 


 \subsubsection{Flow-field measurements: Stereoscopic time-resolved particle image velocimetry}
\label{subsubSec:PIV}
 To further understand the flow, and its relationship to the radiated sound, an experimental study of the aerodynamic field in the interaction region has been carried out. The three components of the velocity field ($u_1$-stream-wise, $u_2$-parallel to the cylinder's span, $u_3$-parallel to the airfoil span) around the airfoil were measured using stereoscopic time-resolved particle image velocimetry (sTR-PIV). LaVision hardware was used to acquire the measurements, while the velocity field was obtained after post-processing on Davis (LaVision-Software). The time between two laser expositions was chosen as $dt=40 $ $\mu\text{s}$. An Antari Z $3000$ fog-machine with a heavy-smoke liquid (ALGAM LIGHTING) was used for the seeding. Details on the sTR-PIV set-up are presented in Appendix \ref{App_PIV}.

    Two planes were considered: i)  perpendicular ($y_1-y_2$)  and ii) parallel ($y_1-y_3$) to the airfoil respectively. The origin of both planes lies at the leading edge and mid-span of the airfoil. For the measurements taken in the $ y_1-y_2$ plane  an acquisition frequency ($f_{\text{acq}}$) of 6 kHz was used while the field of view was $600 \times 200 $ $\text{mm}^2$, that is $(N_{y_1},N_{y_2})=(389,136)$ grid-points in Cartesian coordinates. For the measurements taken in the $ y_1-y_3$ plane  the laser sheet lies $3$ mm above the thickest point of the airfoil, the acquisition frequency was set to 6.25 kHz and the field of view is $400 \times 350$ $\text{mm}^2$, that is $(N_{y_1},N_{y_3})=(286,266)$ grid-points in Cartesian coordinates. The sampling duration for both planes was $T_s=4$s that is  $N_t^{(y_1-y_2)}=24,721$, $N_t^{(y_1-y_3)}=20,000$ snapshots in total for the $y_1-y_2$ and $y_1-y_3$ planes respectively. At the same time with the PIV-measurements two analog microphones have been placed symmetrically at $6.9c$ above and below the center of the airfoil [$\vec{x}_{\mathrm{M}_{16},\mathrm{M}_{17}}=(c/2,\pm 6.9c,0)$]. 

 As demonstrated by the acoustic measurements (figure \ref{Interaction_verticalhorizontal}) when the wake of the cylinder interacts with the downstream airfoil, a $10$ dB increase in the sound pressure levels is observed. Therefore, we focus our attention on the fluctuation field in the vicinity of the airfoil. Animations of the fluctuation field are shown in Movie 1, Movie 2 of the supplementary files for the $y_1-y_2$ and $y_1-y_3$ planes respectively.
 
\begin{figure}
    \centering
    
    \begin{subfigure}{\textwidth}
        \centering
        
        \begin{subfigure}{\textwidth}
            \centering
            \includegraphics[width=\textwidth]{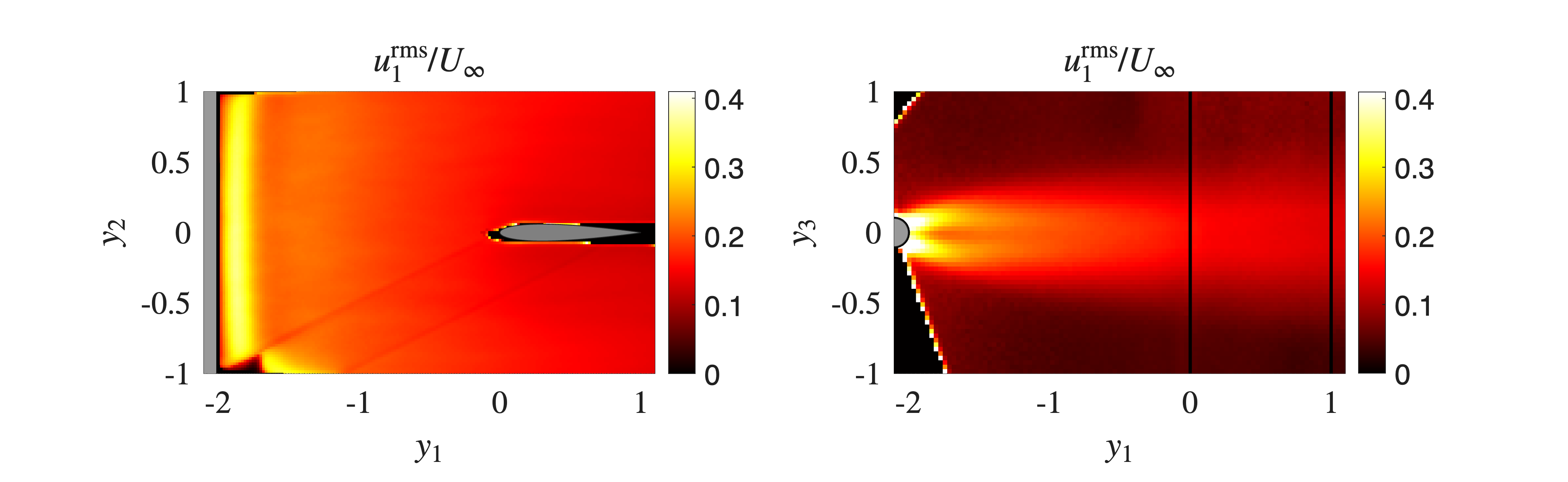}
        \end{subfigure}

        \begin{subfigure}{\textwidth}
            \centering
            \includegraphics[width=\textwidth]{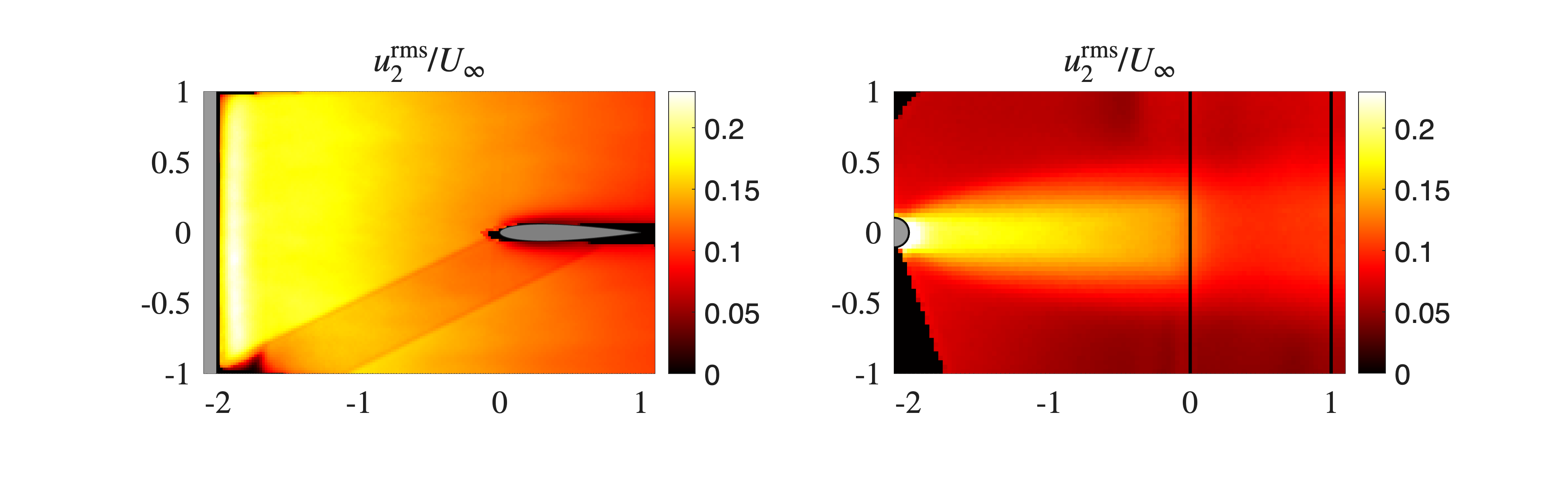}
        \end{subfigure}

        \begin{subfigure}{\textwidth}
            \centering
            \includegraphics[width=\textwidth]{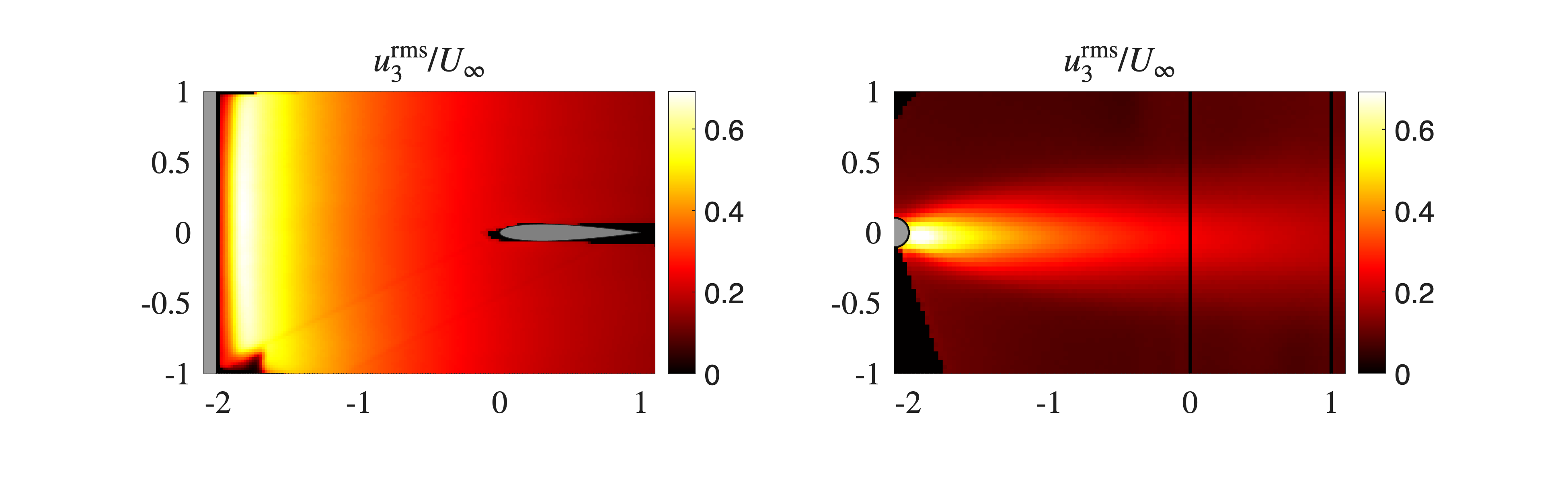}
        \end{subfigure}
  \end{subfigure}
\caption{Normalised root mean square of the velocity-fluctuation field of the three components ($u_1-$top, $u_2$-middle, $u_3$-bottom row) for the $y_1-y_2$ plane ($y_3=0$, left column) and $y_1-y_3$ plane ($y_2=0.03c$, right column). On the $y_1-y_2$ plane the grey area at $y_1/c<-2$ and the thick black line at $y_1/c=-2$ correspond to the cylinder and its boundary respectively. On the $y_1-y_3$ plane (right column of the figure) the vertical lines correspond to the leading and trailing edges of the airfoil.}   \label{u_rms}
\end{figure}

  A measure of the energy of the velocity-fluctuation field is represented by the root mean square (rms) of each fluctuation-velocity component,
 \begin{equation}
u^\text{rms}_i=\sqrt{\frac{1}{N_t}\sum_{n=1}^{N_t}  \left(u'_i \right)^2},
\end{equation}
\noindent where $u'_i$ is the $n-$th snapshot of the $i$-th component of the velocity fluctuation-field. Figure \ref{u_rms} shows the root mean square values of the 3 unsteady velocity-components for both planes, normalised by $U_{\infty}$. Notice that in the two-dimensional problem the cylinder wake is only described by the vortex street which is driven by $u_1',u_3'$, whereas in the present case it can be seen that the velocity fluctuation parallel to the cylinder's span ($u_2'$) arises as well and creates an upwash/downwash velocity field on the leading edge of the airfoil. In contrast to most studies related to cylinder-noise \citep{inoue2002sound,prinja2025experimentally} or the classical rod-airfoil set-up \citep{jacob2005rod}, it will be shown that in the present configuration the sound is generated due to the three-dimensional nature of the flow field and more specifically that the interaction of $u_2'$ with the leading edge of the airfoil generates an acoustic dipole oriented in the $y_2$-direction.

\begin{figure}
  \centering
                \includegraphics[width=0.99\textwidth]{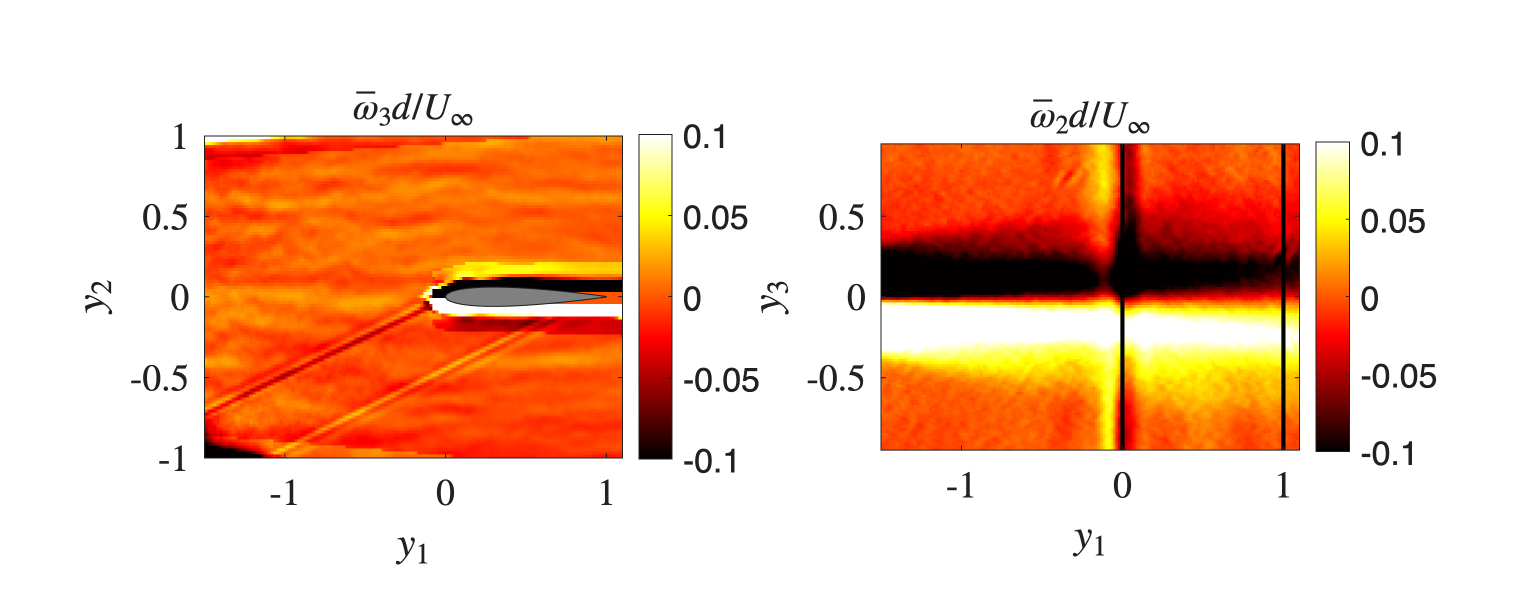}
                \caption{Normalised mean vorticity field for the $y_1-y_2$ plane (left) and $y_1-y_3$ plane (right), non-dimensionalised by the diameter of the cylinder and the flow speed.}
	\label{Mean_vorticity}
    \end{figure}

The vorticity field is obtained by employing a Richardson differentiation scheme (accuracy $O(\Delta x^4)$), which is shown to give accurate results for computing the vorticity field from PIV-data \citep{koschatzky2011study}. The normalised mean vorticity field is shown on figure \ref{Mean_vorticity}.
The normalised mean vorticity component aligned with the airfoil span takes its maximum values close to the boundaries of the airfoil in the $y_1-y_2$ plane and is negligible elsewhere, while the mean vorticity component in the $y_1-y_3$ plane shows that there exist two rows of counter-rotating vorticity symmetrically above and below the cylinder taking their maximum values in a range $y_3 \in [-0.5,0.5]$. However, information about the boundary layer cannot be obtained by the current set-up and the vorticity very close to the rigid airfoil cannot be trusted since it could be an artefact of the differentiation. 

\begin{figure}
  \centering
 \includegraphics[width=0.99\textwidth]{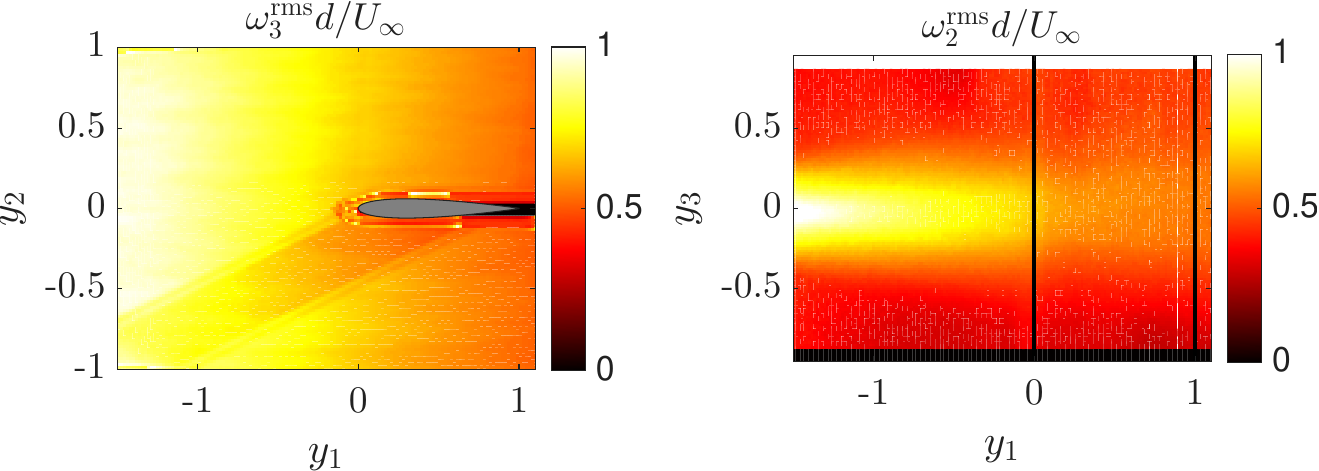}
                \caption{Normalised root mean square of the vorticity-fluctuation components, non-dimensionalised with the diameter of the cylinder and flow speed: i) $\omega_3'$ on the $y_1-y_2$ plane (left) and $\omega_2'$ on the $y_1-y_3$ plane (right).}
	\label{vorticity_rms}
    \end{figure}

The fluctuating vorticity components $\omega_2', \omega_3'$ related to the vortex shedding and the unsteady lift on the airfoil respectively are obtained from the velocity fluctuation field on both planes.  Figure \ref{vorticity_rms} shows the rms values of the vorticity fluctuation fields. It is observed that $\omega_3'$ takes its highest values close to the cylinder and decays further downstream, however its amplitude remains high around the leading edge. It is noted that both components of the rms-vorticity field are by an order of magnitude greater than the respective time-averaged vorticity components and therefore we are permitted to assume that, 
$$\omega_{2,3}^{\mathrm{rms}} \gg  \bar{\omega}_{2,3}.$$
\noindent  The latter observation will be used in following sections, where the modelling of the aeroacoustic source is considered.

To summarise, the three-dimensional nature of the flow field leads to an acoustic dipole oriented along the cylinder's span, in contrast to the purely two-dimensional problem. In the following, we seek the link between the flow-field and the observed interaction noise. 

\subsection{Link between acoustics and flow field}
\label{subsubSection:Combined}
The latter can be explored by computing the coherence of the acoustic and flow fields. It is acknowledged that coherence does not necessarily imply a cause and effect relationship between the two quantities, however in the present work it allows us to form hypotheses about the links between the flow and sound fields that will be later examined via the models of Section \ref{sec:semi_empirical_models}. Microphones $16, 17$ (figure \ref{Set_up_with_mics}) are synchronised with the flow measurements taken in both planes.

\begin{figure}
\centering
   \includegraphics[width=\textwidth]{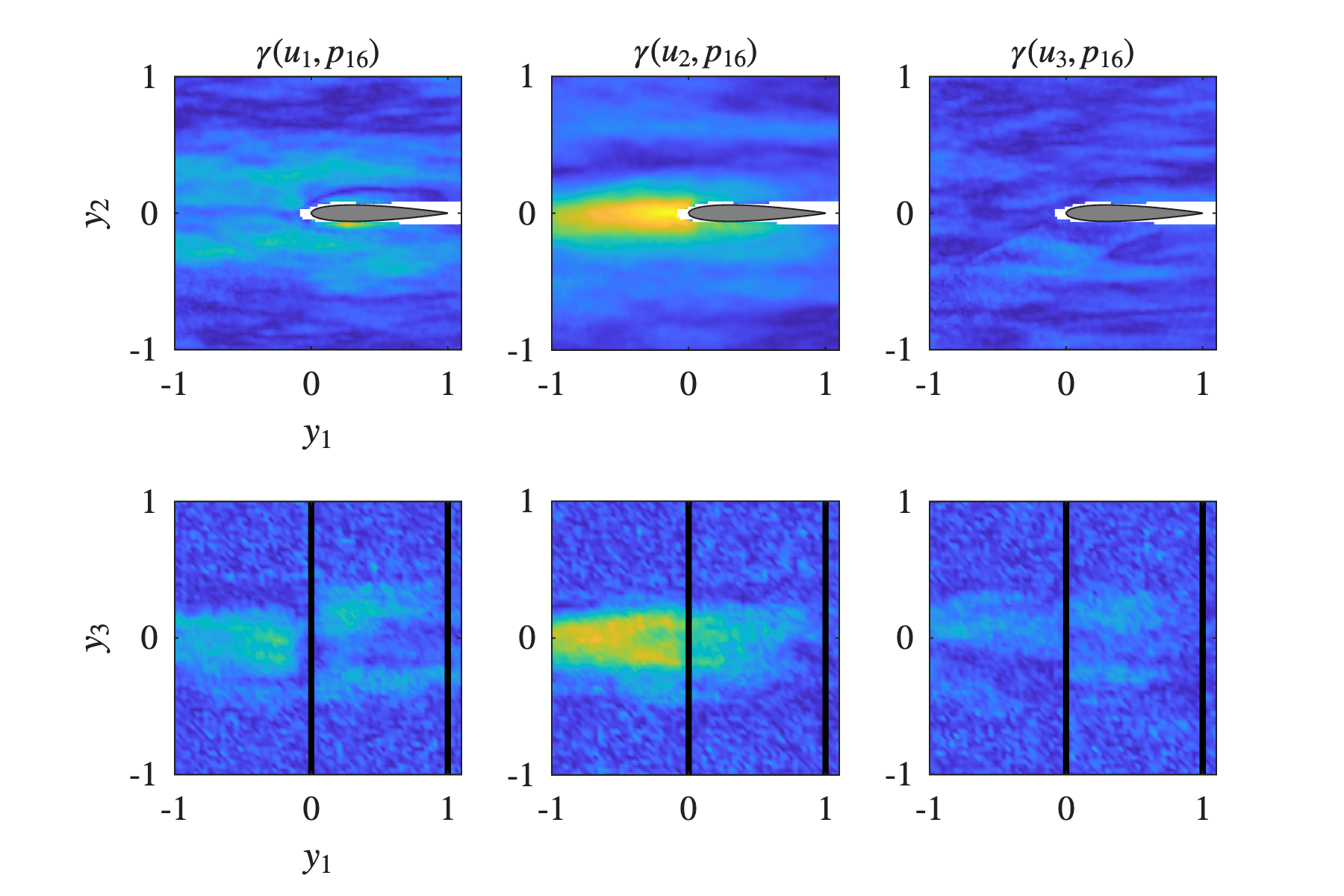}
    \caption{Coherence plots between $p_{16}$ and the three velocity fluctuation components: $u'_1$ (top row), $u'_2$ (middle row), $u'_3$ (bottom row), for the $y_1-y_2$ plane (left column) and $y_1-y_3$ plane (right column), at $\mathrm{St}=0.38$. The solid black vertical lines correspond to the leading and trailing edges of the airfoil. The colormap is kept the same in all figures and takes values from 0 (blue) to 0.6 (yellow).}
\label{coherence_rod_airfoil_St038}
\end{figure}

  The coherence between the acoustic and the velocity fields is computed as
\begin{equation}
    \gamma_{ab}(St)= \frac{\left| S_{ab}(St) \right|}{\sqrt{S_{aa}(St)S_{bb}(St)}},
    \label{coherence_function}
\end{equation}
where $a,b$ denote the signals for which the coherence is computed, $S_{ab}$ is the cross spectral density between $a,b$ and $S_{aa}, S_{bb}$ are the power spectral densities of signals $a,b$ respectively. The cross- and power-spectral densities were obtained by employing Welch's method with $N_\text{fft}=256$ number of points in each block, $50 \%$ overlap and a Hanning window. The coherence plots between the three velocity components and the microphone located above the airfoil (Mic. 16) are shown in figure \ref{coherence_rod_airfoil_St038} for the frequency of the radiation peak $(\mathrm{St}=0.38)$ and both planes ($y_1-y_2$; $y_1-y_3$). It is observed that the unsteady vertical velocity fluctuation at the leading edge ($u_2'$) shows the maximum coherence ($\gamma_{u_2'p_{16}} =0.6$) with the acoustic field (middle row of figure \ref{coherence_rod_airfoil_St038}) compared to the streamwise and airfoil-spanwise velocity fluctuation components $\mathrm{St}=0.38$ (first and last row of figure \ref{coherence_rod_airfoil_St038}).

It can be therefore presumed that the interaction-noise is associated with the unsteady lift generated by the wake-airfoil interaction. The frequency band of this flow-acoustic relationship can be estimated by computing the coherence of the $u_2'$-velocity component at different frequencies.

\begin{figure}
\centering
   \includegraphics[width=1.00\textwidth]{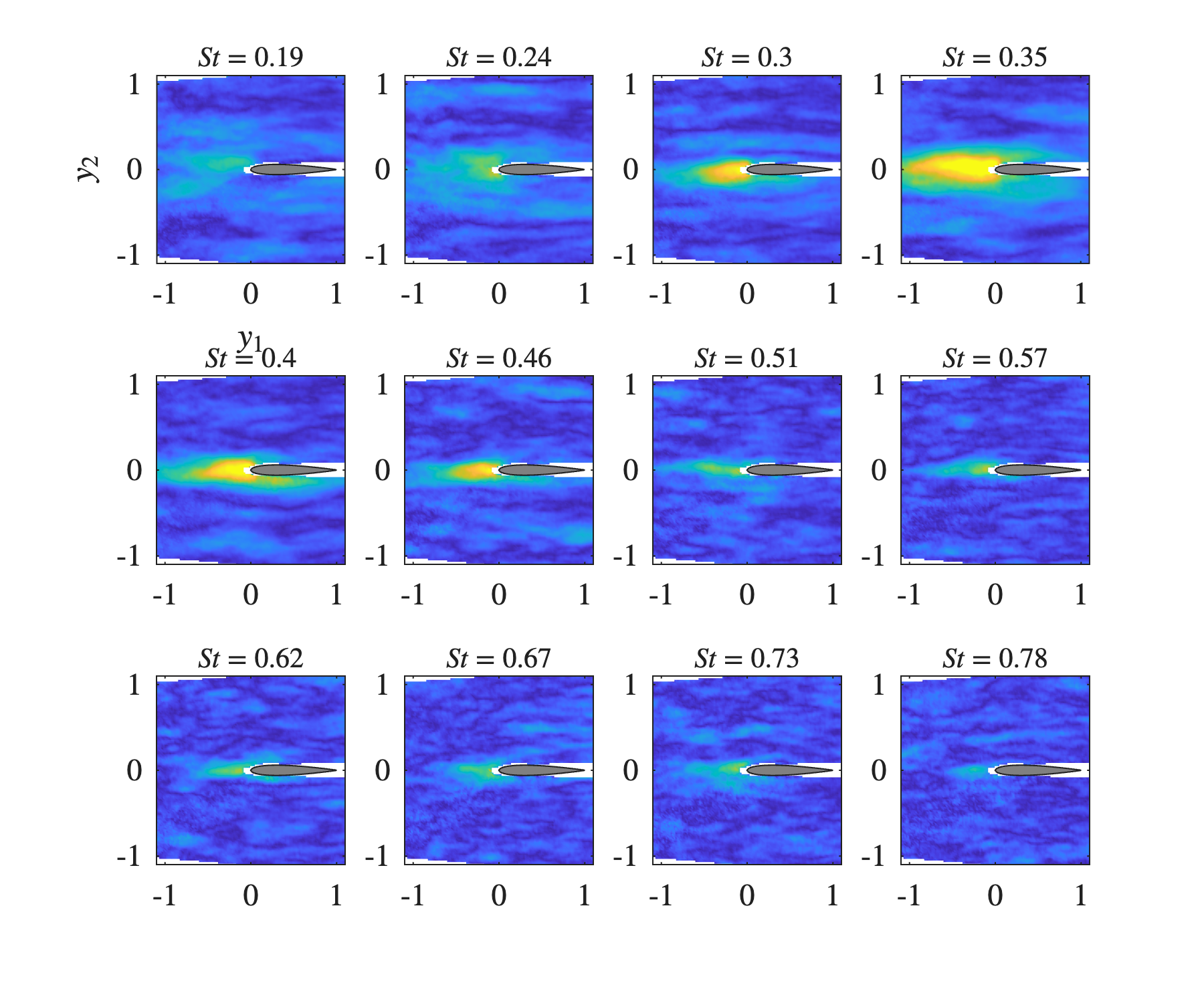}
    \caption{Coherence of the upwash velocity and the microphone located $6.9c$ above the airfoil $(\gamma_{u'_2p_{16}})$ for a range of Strouhal numbers. The colormap is kept the same in all figures and takes values from 0 (blue) to 0.6 (yellow).}
\label{Coherence_frequency_band}
\end{figure}

  Figure \ref{Coherence_frequency_band} shows the coherence plots of the upwash/downwash velocity at the leading edge with Mic. 16 for $\mathrm{St}=0.19$ up to $\mathrm{St}=0.78$. The maximum coherence is observed at the double of the vortex shedding frequency ($\mathrm{St}=0.38$), while it decays for higher Strouhal numbers. The latter indicates that the interaction-noise, produced when the orthogonally-placed cylinder's wake interacts with the downstream airfoil is extended over a larger frequency-band, while the maximum coherence values are found for $\mathrm{St} \in [0.3,0.46].$ . 

Once the flow-quantities linked to the sound generation mechanisms are identified, in the present case the upwash/downwash unsteady velocity ($u_2$) near the leading edge, we seek a simplified representation of the latter to extract the acoustically important coherent structures of the flow field that will be achieved via a vortex-sound model and further processing of the data as discussed in sections \ref{sec:TVS}, \ref{sec:semi_empirical_models}.

\subsection{Summary of experimental results}
\label{subSec:Discussion_Experiments}
An experimental study of the aeroacoustics of a perpendicular cylinder's wake-airfoil interaction has been conducted where acoustic and velocity measurements (sTR-PIV) were obtained to characterise the sound field and its link with the unsteady aerodynamics. The experimental observations indicate that the sound radiation is composed of two dipoles: i) a dipole oriented parallel to the unsteady lift acting on the cylinder ($y_3$-direction), that mainly captures the cylinder-noise and ii) a dipole oriented parallel to the direction of the unsteady lift acting on the airfoil, that is an outcome of the wake-airfoil interaction. 

The coherence maps between the acoustic and flow-fields showed that the unsteady, cylinder-span aligned velocity component at the leading edge dominates the observed wake-airfoil interaction noise. The frequency range where the maximum coherence of $u_2$ with the acoustic signal, measured by the microphone above the center of the airfoil, and maximum sound pressure levels occur is $\mathrm{St} \in [0.3,0.46].$

The latter observations in combination with the acoustically important coherent structures shown in \cite[figure 14]{do_amaral_perpendicular_2025} lead to the conclusion that the von Kármán vortices shed in the wake are distorted, possibly by a secondary instability, giving rise to a waviness aligned with the cylinder-span that results in vorticity fluctuations aligned with the airfoil span, thus imposing an upwash/downwash fluctuation field which is presumed to drive the interaction noise. The latter would not appear in a purely two-dimensional problem where $\omega_2$-vorticity interacts with the airfoil's leading edge \citep{schlinker1983rotor,ahmadi1986experimental,amiet1986airfoil,howe1988contributions, howe1989unsteady}. In what follows, a simplified, linearised model will be used to back the previous assumption and shed light to the flow structures linked to sound production mechanisms.

\section{Vortex-sound formulation}
\label{sec:TVS}
The experimental campaign has shown that the aeroacoustic source consists of $\omega_3-$vorticity that impinges on the leading edge of the airfoil. Those vorticity fluctuations are scattered by the leading edge and an acoustic dipole is produced oriented in the direction of the unsteady lift acting on the airfoil. Neglecting distortion of the incident vorticity field by the airfoil, we use the incident airfoil-span-aligned vorticity field to describe the source term. The aim of the present section is to carry out a more in depth analysis regarding the physics of the problem, to propose a simplified, experimentally-informed model and via this model to investigate the key parameters related to sound-generation. 

Following what was shown, as the vorticity field is used to represent the coherent-structures of the flow that are correlated to the acoustic-field, Powell-Howe's vortex-sound theory \citep{powell1964theory, howe1975contributions} is used as the basis of the modelling. A  simplified three-dimensional model of the acoustic-pressure field will be provided by linearising the Lamb-vector and using Howe's compact Green's functions. 

\subsection{Linearisation and simplification of the Lamb-Vector }
\label{subSec:TVS}
 For low Mach numbers, the acoustic pressure in the frequency domain is given (in dimensional units) by \citet{howe1975contributions} as
\begin{equation}
\hat  p=-\rho \int_{\tilde V} \left( \vec{  \tilde  \omega} \times \vec{  \tilde  u} \right) \cdot \vec{\tilde \nabla}  \tilde G  \mathrm{d}  \tilde V,
     \label{TVS_f_dim_unit}
\end{equation}
\noindent We non-dimensionalise the flow quantities  (vorticity, velocity) according to section \ref{subSec:Measurements}, the gradient of the Green function $\vec{\tilde \nabla}  \tilde G=\frac{1}{c^2} \vec \nabla G,$ and following this change of variables we integrate on the non-dimensional grid $\tilde V = c^3 V$. Thus, equation \eqref{TVS_f_dim_unit} reads as,
\begin{equation}
 \hat p=-\rho U_\infty^2 \frac{c}{d} \int_V \left( \vec{   \omega} \times \vec{   u} \right) \cdot \vec{\nabla}    G \mathrm{d} V,
     \label{TVS_f}
\end{equation}
where $\vec{\omega} \times \vec{u}$ is the Lamb-vector, $ \vec{\nabla}=(\frac{\partial }{\partial y_1},\frac{\partial }{\partial y_2},\frac{\partial}{\partial y_3})$, $G$ the Green function, tailored to the airfoil . $V$ corresponds to the integration region, normalised by the chord length $\left(c\right)$, which is defined as the region where $\vec{\omega} \neq 0$ and $\rho$ to the fluid density in the propagation medium at rest, assumed constant. Notice that $\hat p$ is kept in its dimensional units to facilitate comparisons with the acoustic measurements. By decomposing the flow field in mean and fluctuating quantities, the Lamb vector reads as,
\begin{equation}
   \begin{aligned}
    \vec{\omega} \times \vec{u} = &\vec{e}_1 \left[\left(\bar{\omega}_2 + \omega_2' \right) \left(\bar{u}_3+ u_3' \right)  -  \left(\bar{\omega}_3 + \omega_3' \right) \left(\bar{u}_2+ u_2' \right) \right] - \\
    &\vec{e}_2 \left[\left(\bar{\omega}_1 + \omega_1' \right) \left(\bar{u}_3+ u_3' \right)  -  \left(\bar{\omega}_3 + \omega_3' \right) \left(\bar{u}_1+ u_1' \right) \right] + \\ 
    &\vec{e}_3 \left[\left(\bar{\omega}_1 + \omega_1' \right) \left(\bar{u}_2+ u_2' \right)  -  \left(\bar{\omega}_2 + \omega_2' \right) \left(\bar{u}_1+ u_1' \right) \right] 
    \end{aligned}
     \label{Lamb2}
\end{equation}
Since the problem will be treated in the frequency domain all the fluctuation quantities will be replaced with their frequency-domain counterparts, 
\begin{equation*}
    q_i' \to \hat{q}_i.
\end{equation*}
Combining Equations (\ref{TVS_f}), (\ref{Lamb2}) the acoustic field can be expressed as the sum of streamwise ($\hat p_{y_1}$), cylinder-span ($\hat p_{y_2}$) and airfoil-span ($\hat p_{y_3}$) oriented dipoles 
\begin{equation}
   \begin{aligned}
 \hat p &= \hat p_{y_1} +\hat p_{y_2} +\hat p_{y_3} \\
  \hat     p_{y_1} &= -\rho  U_\infty^2 \frac{c}{d}\int_V    \left[\left(\bar{\omega}_2 + \hat \omega_2 \right) \left(\bar{u}_3+ \hat  u_3 \right)  -  \left(\bar{\omega}_3 + \hat  \omega_3 \right) \left(\bar{u}_2+ \hat u_2 \right) \right]  \frac{\partial G}{\partial y_1} \mathrm{d}V \\
 \hat     p_{y_2} &= \rho U_\infty^2 \frac{c}{d}\int_V \left[\left(\bar{\omega}_1 + \hat  \omega_1 \right) \left(\bar{u}_3+ \hat u_3 \right)  -  \left(\bar{\omega}_3 + \hat \omega_3 \right) \left(\bar{u}_1+ \hat  u_1 \right) \right]  \frac{\partial G}{\partial y_2} \mathrm{d}V \\
 \hat    p_{y_3}&= - \rho U_\infty^2 \frac{c}{d} \int_V  \left[\left(\bar{\omega}_1 + \hat  \omega_1 \right) \left(\bar{u}_2+ \hat u_2\right)  -  \left(\bar{\omega}_2 + \hat  \omega_2 \right) \left(\bar{u}_1+ \hat  u_1 \right) \right] \frac{\partial G}{\partial y_3} \mathrm{d}V
    \end{aligned}
     \label{Dipoles}
\end{equation}
\noindent The observed interaction-noise is associated with the cylinder-span oriented dipole ($\hat p_{y_2}$) corresponding to the second term of Equations (\ref{Dipoles}). From now on we will consider $\hat p \equiv \hat p_{y_2}$ and it follows,
\begin{equation}
   \begin{aligned}
 \hat   p= \rho U_\infty^2 \frac{c}{d} \int_V \left[\left(\bar{\omega}_1 \bar{u}_3+ \hat   \omega_1 \bar{u}_3 +\bar{\omega}_1  \hat  u_3 + \hat  \omega_1 \hat  u_3\right)   -  \left(\bar{\omega}_3 \bar{u}_1  + \hat \omega_3 \bar{u}_1+\bar{\omega}_3 \hat u_1 + \hat  \omega_3 \hat  u_1\right) \right]  \frac{\partial G}{\partial y_2} \mathrm{d}V.
    \end{aligned}
     \label{Dipoles2}
\end{equation}

Equation \eqref{Dipoles2} is further simplified by considering the experimental data. From the flow-measurements we obtain $|\bar{u}_3| \approx  |\bar{u}_2|\ll |\bar{u}_1|$ close to the wake/airfoil interaction-region and hence the product of fluctuation quantites with $\bar{u}_3$ can be neglected to a first order approximation. Then, it follows
\begin{equation}
   \begin{aligned}
    \hat p= \rho U_\infty^2 \frac{c}{d} \int_V \left[\left( \bar{\omega}_1  \hat  u_3 + \hat  \omega_1 \hat  u_3\right)   -  \left(\bar{\omega}_3 \bar{u}_1  +\hat   \omega_3  \bar{u}_1+\bar{\omega}_3 \hat  u_1 + \hat  \omega_3 \hat u_1\right) \right]  \frac{\partial G}{\partial y_2} \mathrm{d}V.
    \end{aligned}
     \label{Dipoles3}
\end{equation}

\noindent In Appendix \ref{Lamb} we demonstrate that to a first order approximation $$\left( \bar{\omega}_1  \hat  u_3 + \hat  \omega_1 \hat  u_3\right)   -  \left(\bar{\omega}_3 \bar{u}_1  +\hat   \omega_3  \bar{u}_1+\bar{\omega}_3 \hat  u_1 + \hat  \omega_3 \hat u_1\right) \approx \bar{u}_1 \hat{\omega}_3 .$$ \noindent Therefore,  \eqref{Dipoles3} becomes
\begin{equation}
\begin{aligned}
    \hat p  =-\rho U_\infty^2 \frac{c}{d}  \int_V \bar{u}_1 \hat{\omega}_3  \frac{\partial G}{\partial y_2} \mathrm{d}V.
    \end{aligned}
     \label{TVS_simplified}
\end{equation}

\subsection{Acoustic scattering: the tailored Green function}
\label{subSection:GreenFunc}
The Green function is used to account for the acoustic scattering of the source-term by the airfoil. 
The  frequency band of interest $\mathrm{St} \in [0.2, 0.46] $  corresponds to non-dimensional acoustic wavenumbers 
\begin{equation}
k=2\pi  \frac{c}{d} \mathrm{St} \mathrm{M}.
\label{eq:ac_wvnbr}
\end{equation}

\noindent For an airfoil of chord $c=100$ mm these non-dimensional wavenumbers correspond to wavelengths 
\begin{align*}
\lambda = \frac{2 \pi}{k} c \geq 5 c.
\end{align*}

 \noindent Thus, we are allowed to apply a low-frequency approximation of the Green function for the present problem.

In particular, we use Howe's theory of the compact Green functions to describe the scattering of an acoustic wave impinging on an obstacle of arbitrary geometry and characteristic dimension (here $c$). For observers located many wavelengths far from the aerodynamic source and the rigid geometry immersed in the flow, \citet{Howe_1975} has demonstrated that the Green function can be calculated by using the reciprocal theorem as

\begin{equation}
G \left(\vec{x};\vec{Y},\mathrm{St} \right)= \frac{- \exp \left( i k \left| \vec{x}-\vec{Y} \right| \right)}{4 \pi \left| \vec{x}-\vec{Y} \right| },
 \label{CompactGF}
 \end{equation}
where $k = 2\pi c/ \lambda$ the non-dimensional wavenumber of the sound wave of eq. \eqref{eq:ac_wvnbr}. 
The Kirchhoff vector, 
$$ \vec{Y}=\vec{y}-\phi^*(\vec{y}),$$ corresponds to the position vector of the source corrected by a potential field $\phi^*(\vec{y})$ that accounts for the distortion of the flow field due to the existence of a rigid obstacle. The Kirchhoff-vector's components ($Y_j$) are defined as the velocity potential of a uniform flow with unit velocity in the j-th direction and are obtained by solving the Laplace equation with associated boundary conditions \citep{Howe2002TheoryOV}, 

    \begin{equation}
        \begin{aligned}
            \nabla^2Y_j &=0\\
            \frac{\partial Y_j}{\partial y_n} &=0\\
            Y_j & \sim y_j \> \> \text{as} \> \> |\vec{y}|\to \infty,
        \end{aligned}
        \label{pot_flow}
    \end{equation}
 \noindent   where $\frac{\partial Y_j}{\partial y_n}$ denotes the normal derivative on the boundary. Notice that the Kirchhoff vector is not related to the actual flow field, it is rather used to compute the acoustically compact Green function. Setting $\phi^*(\vec{y})=0$, we retrieve the free-field Green function. The Kirchhoff vector $\vec{Y}$ is given in its non-dimensional form (normalised by the chord-length). For an observer in the far field, we apply the following approximations \citep{Howe2002TheoryOV} :
\begin{equation}
\begin{aligned}
    |\vec{x}-\vec{Y}| & \approx |\vec{x}|-\frac{\vec{x} \cdot \vec{Y}}{|\vec{x}|}, \\ 
    \frac{1}{|\vec{x}-\vec{Y}|} & \approx \frac{1}{|\vec{x}|}.
    \end{aligned}
\end{equation}

\noindent Taking the first spatial derivative in the $j$-th direction we obtain : 
\begin{equation}
    \frac{\partial G}{\partial y_j} \left(\vec{y},\vec{x},\mathrm{St} \right)=-\mathrm{i}k\frac{e^{\mathrm{i}k|\vec{x}|} e^{-\mathrm{i}k \frac{\vec{x} \cdot \vec{Y}}{|\vec{x}|}}}{4\pi|\vec{x}|} \left( \frac{x_1}{|\vec{x}|} \frac{\partial Y_1}{\partial y_j}+ \frac{x_2}{|\vec{x}|} \frac{\partial Y_2}{\partial y_j}+\frac{x_3}{|\vec{x}|} \frac{\partial Y_3}{\partial y_j} \right),
    \label{newGF}
\end{equation}
The previous expression is used extensively to model problems of sound generation by fluid-structure interactions \citep{kambe1985acoustic,howe_1998,Howe2002TheoryOV,spiropoulos_aeroacoustics_2025}.

The Kirchhoff vectors can be obtained by solving eq. (\ref{pot_flow}). Analytical solutions exist for simple shapes such as spheres, cylinders, rigid strips, semi-infinite half-planes or wedges and others \citep{Howe2002TheoryOV} . However, taking into account the actual geometry of an airfoil or a more complicated shape requires numerical solutions of eq. (\ref{pot_flow}) or other techniques based on complex analysis. For instance, \citet{crowdy2010new} developed a methodology to calculate the potential flow around multiple rigid objects  immersed in a flow which later was applied by \citet{baddoo2019compact} to demonstrate how compact Green functions can be constructed by applying this methodology. Other studies use a numerical implementation of the Schwartz-Christoffel mapping, to obtain the Kirchhoff vector by transforming the physical space into the complex plane, solving the potential flow equation and then returning to the physical space \citep{HARWOOD2016795}.

In the present work we are interested in modelling finite, two-dimensional shapes with an infinite span. Taking advantage of the homogeneous span-wise direction we may write \citep{Howe2002TheoryOV} , $$Y_3\approx y_3.$$ A simple and efficient way to obtain numerical solutions of eq. (\ref{pot_flow}) is to employ the source panel-method \citep{anderson2011ebook,fletcher2012computational}. The surface of the body is divided into panel segments with a corresponding distribution of sources and the uniform flow is parallel to the $y_1$ or $y_2$ axis. The algorithm of the panel method code is adapted from \citet[p.130-136]{fletcher2012computational}. As a result, the Kirchhoff vector components $Y_j$ are computed by introducing a free-field velocity $U_j=1$ in the $j$-th direction \citep{Howe2002TheoryOV} . The derivatives of the Kirchhoff vector components correspond to the velocity field obtained by the potential flow solution, that is $\frac{\partial Y_1}{\partial y_2}$ corresponds to the $y_2$-aligned velocity field obtained by the complex potential $Y_1$ for $U_1=1$ and similarly $\frac{\partial Y_2}{\partial y_2}$ is computed by the $y_2$-aligned velocity field obtained by the complex potential $Y_2$ for $U_2=1$. We used $900$ panels for the NACA-0012 surface mesh, in order to assure the convergence of the method. Appendix \ref{panel} shows a validation of the numerical method.

Combining Equations  (\ref{TVS_simplified}) and (\ref{newGF}) it follows,
\begin{equation}
  \hat p\left( \vec{x},\mathrm{St} \right)=- \rho U_\infty^2 \frac{c}{d} \iiint_{V} \bar{u}_1  \left( y_1, y_2\right) \hat \omega_3 \left( \vec{y}, \mathrm{St} \right) \frac{\partial G}{\partial y_2} \left( \vec{y}, \mathrm{St} \right) \mathrm{d} y_1 \mathrm{d} y_2 \mathrm{d} y_3 ,
    \label{model}
\end{equation}
 where from eq. (\ref{newGF}) we obtain 
 \begin{equation}
    \frac{\partial G}{\partial y_2} \left(\vec{y},\vec{x},\mathrm{St} \right)=-\mathrm{i}k\frac{e^{\mathrm{i}k|\vec{x}|} e^{-\mathrm{i}k \frac{\vec{x} \cdot \vec{Y}}{|\vec{x}|}}}{4\pi|\vec{x}|} \left( \frac{x_1}{|\vec{x}|} \frac{\partial Y_1}{\partial y_2}+ \frac{x_2}{|\vec{x}|} \frac{\partial Y_2}{\partial y_2} \right),
    	\label{Eq:Kirchhoff}
\end{equation}

 \noindent Equation (\ref{model}) needs to be informed with the mean velocity and the vorticity fluctuation fields.

\subsection{Generating source fields from PIV data}
\label{subSection:Datamodel}
\begin{figure}
\centering
   \includegraphics[width=0.75\textwidth]{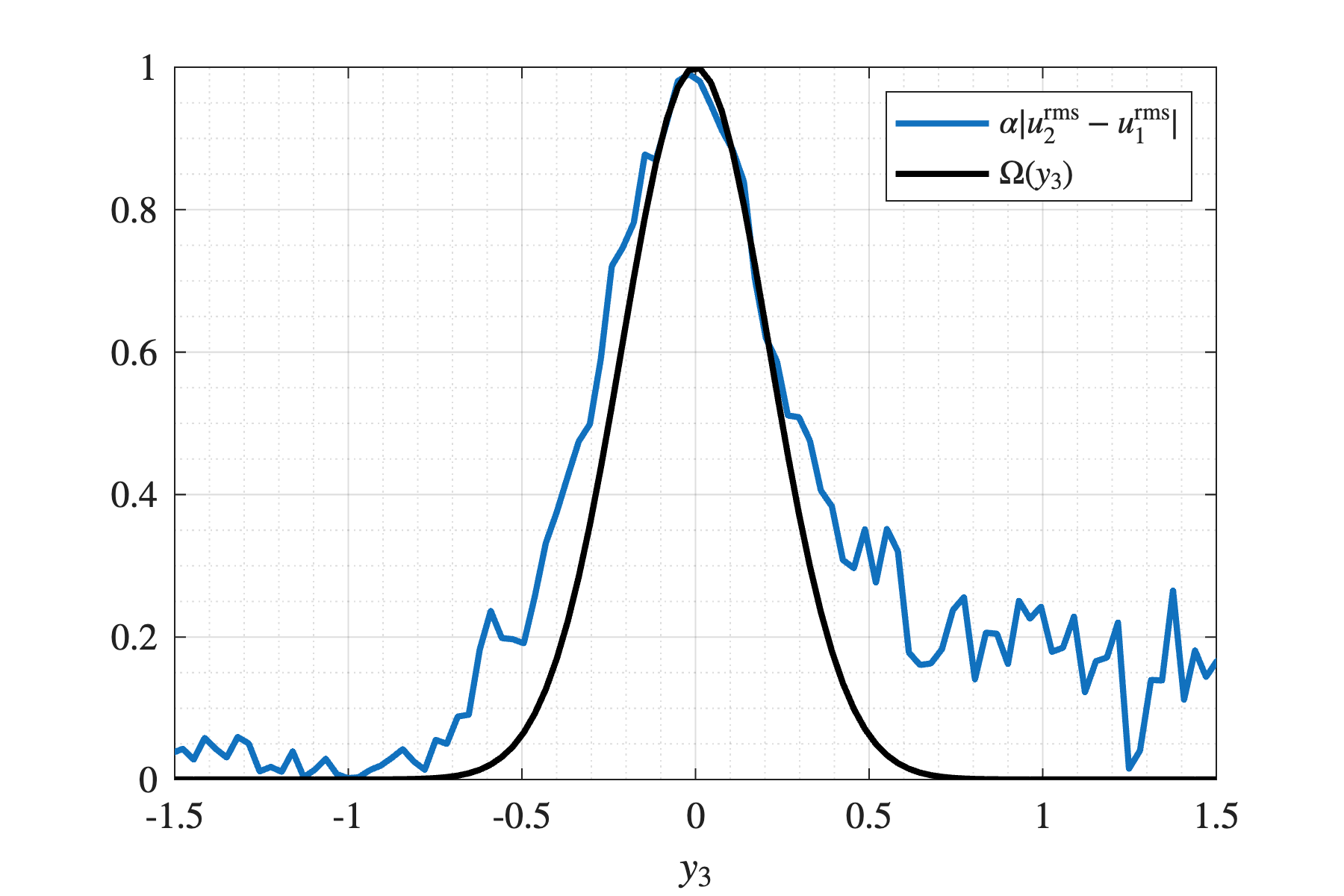}
    \caption{Experimentally fitted support of the airfoil-span-aligned vorticity fluctuations at the leading edge. }
\label{Gaussian_y3}
\end{figure}

In what follows, an estimation of the sound field is provided by eq. (\ref{model}) using the experimental data as input. In reality, the vorticity field is three-dimensional, $\hat{\omega}_3(\vec{y},St)$, however, the experimental set-up does not provide information on the distribution of the airfoil-span-aligned vorticity component along the airfoil's span ($y_3$). To proceed with the modelling we assume that the non-dimensional vorticity component $\left( \hat \omega_3(\vec{y},St) \right)$ can be written as 
\begin{equation}
   \hat  \omega_3\left(\vec{y},\mathrm{St} \right)=\hat \omega_3^{\text{2-D}}(y_1,y_2,St) \Omega_3(y_3),
    \label{Vorticity_2d_3d}
\end{equation}

\noindent where $\hat \omega_3^{\text{2-D}}(y_1,y_2,St) $ corresponds to the reconstructed vorticity field, computed by the Fourier-transformed fluctuation-velocity components ($\hat{u}_1,\hat{u}_2$) of the PIV-data in the plane $y_3=0$ and $ \Omega_3(y_3)$ is a non-dimensional distribution function along the airfoil's span that needs to be specified. The distribution  $ \Omega_3(y_3)$ is not readily available from the experimental data and therefore, we inform it by considering the extent of $u_1^{\mathrm{rms}},u_2^{\mathrm{rms}}$ in the $y_3-$ direction. The airfoil-span aligned vorticity is given by
$$\hat \omega_3 = \frac{\partial \hat u_2}{\partial y_1}- \frac{\partial \hat u_1}{\partial y_2},$$
and since the operators $\partial/ \partial y_1, \> \> \partial/\partial y_2$ do not influence the $y_3-$dependence we assume that the distribution $\Omega_3(y_3)$ is approximated as 
$$\Omega_3(y_3) \sim \alpha |u_2^{\mathrm{rms}}(y_3)-u_1^{\mathrm{rms}}(y_3)|,$$

\noindent where $\alpha$ a calibration factor ensuring that $\Omega_3$ is unitary at the midspan. The spanwise distribution $|u_2^{\mathrm{rms}}(y_3)-u_1^{\mathrm{rms}}(y_3)|,$ at the leading edge of the airfoil ($y_1 \approx 0$) is fitted by a Gaussian distribution
\begin{equation}
    \Omega_3(y_3) =e^{-\frac{y_3^2}{L_3^2}},
    \label{y3_depndence}
\end{equation}
where  $L_3=1.5d/c$ as shown in figure \ref{Gaussian_y3}.
The fit is considered to capture reasonably the distribution of the fluctuation field along the span of the airfoil. For the remainder of the manuscript it is assumed that the distribution along the span remains as in Equation (\ref{y3_depndence}) for all $\mathrm{St}$. In the following, we then take :

\begin{equation}
  \hat  \omega_3(\vec{y},St)=\hat  \omega_3^{\text{(2-D)}}(y_1,y_2,\mathrm{St}) e^{-y_3^2/ L_3^2}.
    \label{Vorticity_2d_3d2}
\end{equation}

Substituting (\ref{Eq:Kirchhoff}) and (\ref{Vorticity_2d_3d2}) into (\ref{model}), it follows that,

\begin{equation}
\begin{aligned}
     \hat p\left(\vec{x},\mathrm{St} \right) &= \frac{i  \rho  U_\infty^2e^{\mathrm{i}k|\vec{x}|}}{4 \pi  |\vec{x}|}   \frac{x_2}{|\vec{x}|}  \frac{ \mathcal{L}_3(x_3,\mathrm{St})}{d}  \times \\
  &  \iint_{-\infty}^{\infty}  \bar{u}_1 \left(y_1, y_2 \right) \hat \omega_3^{\text{2-D}}(y_1,y_2,\mathrm{St})    \left( \frac{x_1}{|\vec{x}|} \frac{\partial Y_1(y_1,y_2)}{\partial y_2}+ \frac{x_2}{|\vec{x}|} \frac{\partial Y_2(y_1,y_2)}{\partial y_2} \right) e^{-\mathrm{i}k \left(\frac{x_1 Y_1}{|\vec x|}+\frac{x_2 Y_2}{|\vec{x}|}\right)}\mathrm{d}y_1 \mathrm{d}y_2 ,
\end{aligned}
\label{ff_pressure}
\end{equation}
where 
\begin{equation}
\mathcal{L}_3\left(x_3,\mathrm{St} \right) = c \sqrt{\pi} \left(kL_3 \right) \exp \left(-\frac{k^2 L_3^2}{4}  \left(\frac{x_3}{|\vec{x}|} \right)^2 \right).
\label{L_3new}
\end{equation}

Equation \eqref{L_3new} shows the dependence of the acoustic field along the airfoil-spanwise direction (directivity in the $x_1-x_3$ plane). For an observer placed at the mid-span, the acoustic pressure takes its maximum values, while for observers placed further away from the midspan it drops. 

\subsection{Estimation of the acoustic field}
Equation (\ref{ff_pressure}) is a three-dimensional simplified description of the acoustic field that depends on i) a low-frequency, semi-analytical expression of the Green function and ii) a linearised source-model based on the theory of vortex sound. In what follows, an estimation of the acoustic field is proposed, when the source-term is informed by the experimental data. More specifically, the source term is represented by the mean streamwise velocity component ($\bar{u}_1$) and the vorticity fluctuation field in the frequency domain $\hat \omega_3^{\text{2-D}}\left(y_1,y_2,\mathrm{St}\right)$ in the plane $y_3=0$. The double integral of the right-hand side of Equation (\ref{ff_pressure}) is computed numerically in the grid of the PIV-field view. 

To inform the model, we use as input the frequency domain counterpart of the source signal. For an estimation of the acoustic field we perform a Fourier transform of the source term using the Welch method. The temporal source-signal is segmented in blocks of 128 points with $50 \%$ overlap, corresponding to $63$ positive frequencies (using the single-sided spectrum). A Hanning window is applied to avoid spectral leakage.

\begin{figure}
\centering
   \includegraphics[width=0.95\textwidth]{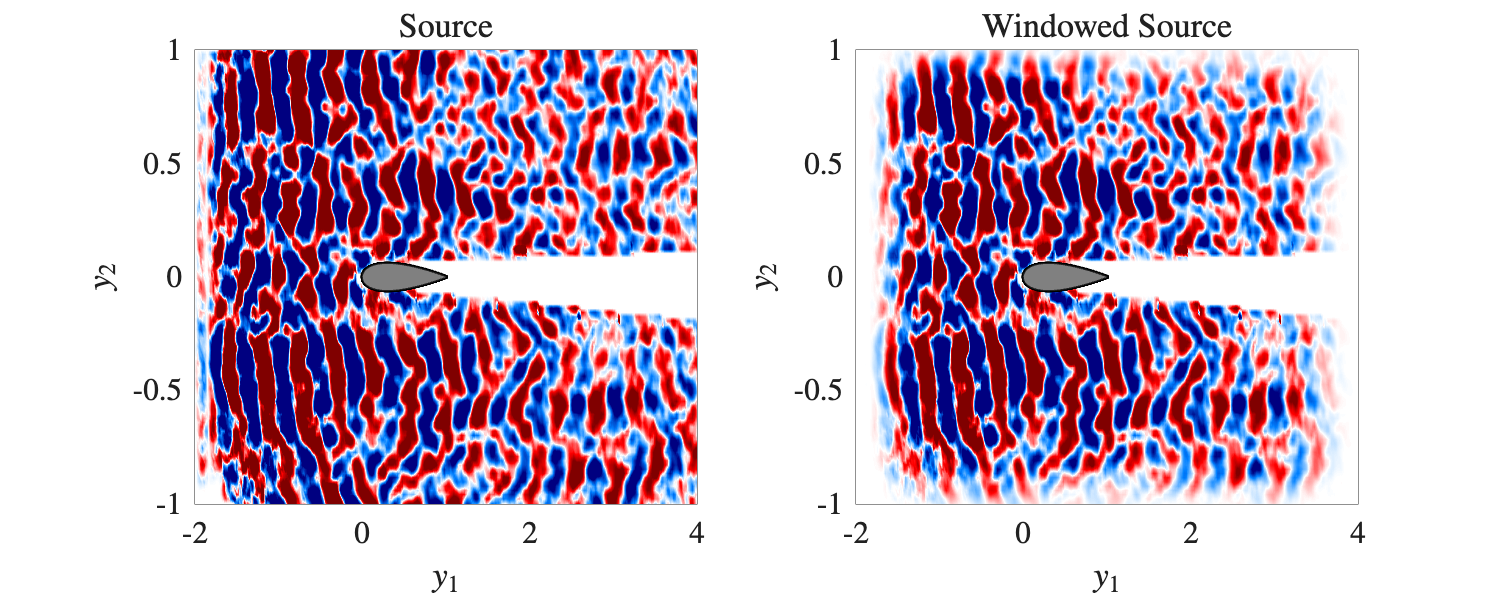}
    \caption{Real part of the linearised source term $\Re[\bar u_1 \omega_3^{\mathrm{2D}}(y_1,y_2,\mathrm{St=0.38}  )]$ vs real part of the windowed linearised source term $\mathcal{W}(y_1,y_2) \Re[\bar u_1 \omega_3^{\mathrm{2D}}(y_1,y_2,\mathrm{St=0.38}  )]$ for $y_{1,c}=1,y_{2,c}=0,L_1=2.45 , L_2=0.90 , n =10$  .}
\label{fig:windowedsource_St038}
\end{figure}

The acoustic field is obtained by performing a numerical integration on the PIV grid. To ensure that the discontinuities at the borders of the integration domain do not affect the computation of the sound field, we use a window function to smooth our data. Following \cite{margnat2014compressibility}, we propose a super-Gaussian function, which reads as
\begin{equation}
	\begin{aligned}
	\mathcal{W}\left(y_1,y_2\right)  &= e^{- \left(\frac{y_1-y_{1,c}}{L_1} \right)^n}e^{- \left(\frac{y_2-y_{2,c}}{L_2} \right)^n},
	\end{aligned}
	\label{eq:super_Gaussian}
\end{equation}
\noindent to smooth the source term at the borders of the domain. The window function lies in $0 \leq \mathcal{W}\left(y_1,y_2\right)  \leq 1$,  $y_{1,c}, y_{2,c}$ are used to center the window function at a specific point in the PIV grid and $L_1,L_2$ correspond to the support of the function along the streamwise and cylinder-span directions accordingly. The exponent $n$ affects the slope of $\mathcal{W}\left(y_1,y_2\right) $ between the values $0,1$. The windowed source term, filtered by a super-Gaussian ellipsoid, is compared to the non-windowed source in Figure \ref{fig:windowedsource_St038} for $y_{1,c}=1,y_{2,c}=0,L_1=2.45 , L_2=0.90 , n =10$ and $\mathrm{St}=0.38$. In what follows, we choose these parameters unless stated otherwise.

Hence for the computation of the acoustic field we write 
\begin{equation}
\begin{aligned}
     \hat p\left(\vec{x},\mathrm{St}\right) &= \frac{i  \rho  U_\infty^2e^{\mathrm{i}k|\vec{x}|}}{4 \pi  |\vec{x}|}   \frac{x_2}{|\vec{x}|}  \frac{ \mathcal{L}_3(x_3,\mathrm{St})}{d}  \times \\
  &  \iint_{\mathrm{PIV-grid}}  \Biggl\{  \bar{u}_1 \left(y_1, y_2 \right) \hat \omega_3^{\text{2-D}}(y_1,y_2,\mathrm{St})  \mathcal{W}(y_1,y_2)  \\
  &\times    \left( \frac{x_1}{|\vec{x}|} \frac{\partial Y_1(y_1,y_2)}{\partial y_2}+ \frac{x_2}{|\vec{x}|} \frac{\partial Y_2(y_1,y_2)}{\partial y_2} \right) e^{-\mathrm{i}k \left(\frac{x_1 Y_1}{|\vec x|}+\frac{x_2 Y_2}{|\vec{x}|}\right)} \Biggr\}\mathrm{d}y_1 \mathrm{d}y_2 ,
\end{aligned}
\label{ff_pressure2}
\end{equation}

The influence of the windowing function on the acoustic field was assessed for different parameters and it was found that it does not influence the source term. A convergence test is shown in Appendix \ref{App:Convergence}.

\begin{figure}
    \centering
    
    \begin{subfigure}{0.75\textwidth}
        \centering
       \includegraphics[width=\textwidth]{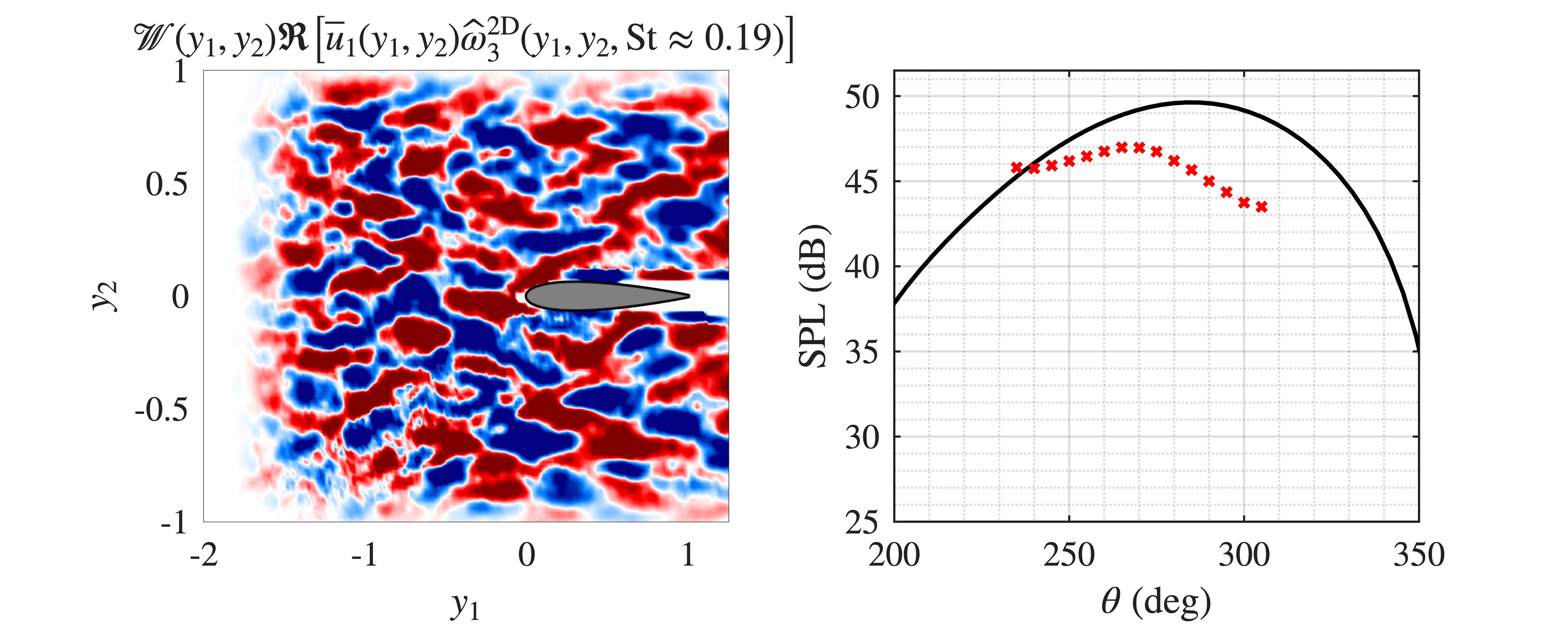}
       \caption{$\mathrm{St}=0.19$}
    \end{subfigure}
    \vspace{0.5cm}
    
    \begin{subfigure}{0.75\textwidth}
        \centering
        \includegraphics[width=\textwidth]{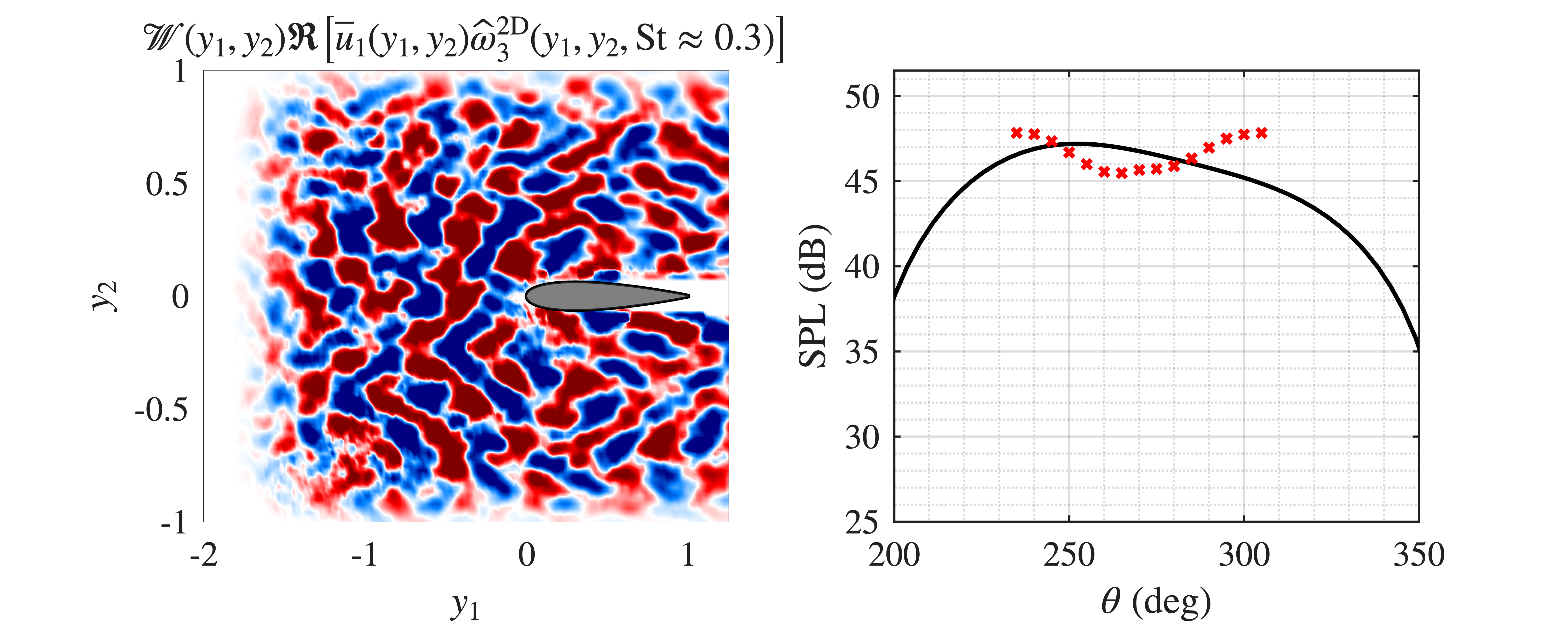}
       \caption{$\mathrm{St}=0.30$}
    \end{subfigure}
    \vspace{0.5cm}

    \begin{subfigure}{0.75\textwidth}
        \centering
        \includegraphics[width=\textwidth]{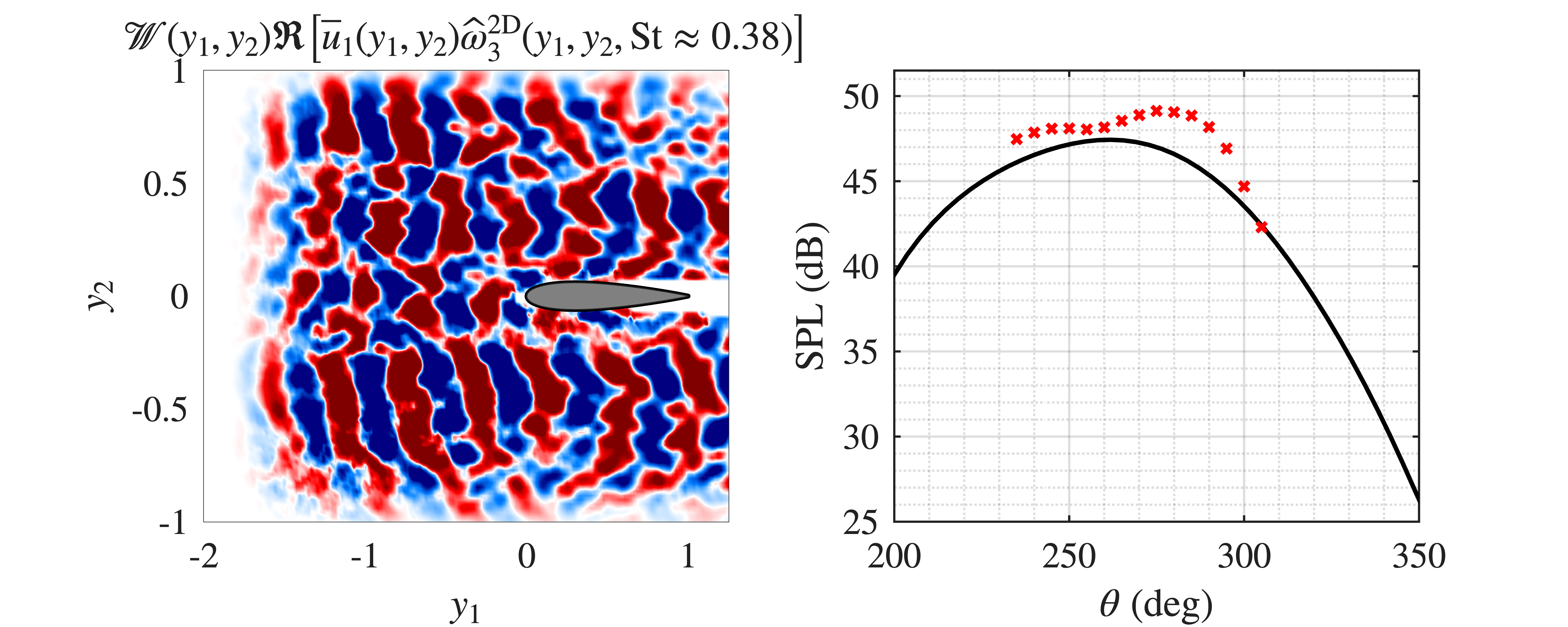}
        \caption{ $\mathrm{St}=0.38$.}
    \end{subfigure}
    \vspace{0.5cm}

    \begin{subfigure}{0.75\textwidth}
        \centering
        \includegraphics[width=\textwidth]{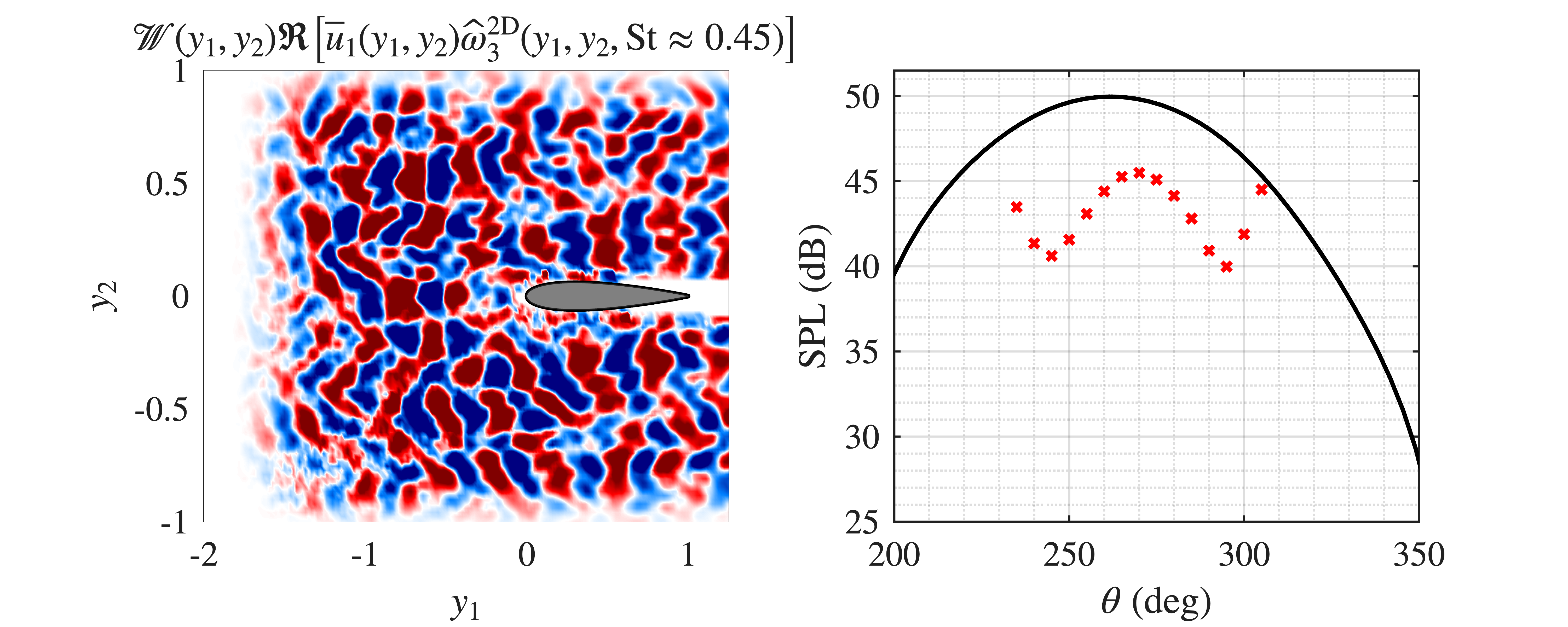}
        \caption{$\mathrm{St}=0.45$. }
    \end{subfigure}

    \caption{Data-based estimation of radiated sound pressure level for different frequencies. The first column corresponds to the real part of the linearised source-term, while the second to a comparison of the computed (black line) vs measured SPL (red crosses) at the corresponding frequency (St).}
    \label{Blocks_fft_data_based_estimation}
\end{figure}

The estimation of the acoustic field (shown in figure \ref{Blocks_fft_data_based_estimation}), based on eq. \eqref{ff_pressure2}, gives an acceptable agreement when compared to experiments for $\mathrm{St}=0.19;0.30;0.38;0.45$. Particularly for $\mathrm{St}=0.38$ where the acoustic radiation peak and maximum coherence are observed, the maximum error between the estimation and the measurements is $2.5$ dB, validating the proposed aeroacoustic model and its underlying assumptions. In particular, the linearised Lamb vector is shown to be an adequate representation of the source-term while the approximate Green function is valid for low frequencies. Notice that for $\mathrm{St}=0.45$, the wavelength becomes comparable to the airfoil's chord $\left( k>1\right)$ and hence the assumption regarding acoustic compactness is no longer valid which leads to a less accurate estimation as shown in figure \ref{Blocks_fft_data_based_estimation}d, that explains the increasing discrepancy between the measurements and eq. \eqref{ff_pressure2}.

\section{Extraction of acoustically important flow-structures}
\label{sec:semi_empirical_models}

The results presented in figure \ref{Blocks_fft_data_based_estimation} indicate that for $\mathrm{St}=0.38$ a clearer spatially-organised structure of the source term is apparent, especially along the streamwise direction where the pronounced waviness is dominated by wavelengths associated with the convective wavenumber $k_1=k/M$.  Therefore, the previous results indicate that the underlying physical mechanisms linked to sound generation at $\mathrm{St}\approx0.38$ can be described in terms of coherent flow-structures of the cylinder's wake. In the rest of the current section we will elaborate on this assumption by further processing the data and introducing a semi-empirical source model. The source distribution is modelled by cylinder-spanwise Fourier modes at $\mathrm{St} = 0.38$ which are convected downstream by $U$ and whose wavenumber and amplitudes are selected from the experimental data.

\subsection{Wavenumber decomposition along the cylinder-span}
\label{subsec:wavenumber_transform}
As previously discussed, the unsteady, linearised Lamb-vector component in $y_2$ gives an adequate representation of the source term and leads to a reasonable computation of the acoustic field when combined with the linearised aeroacoustic model derived in the framework of Howe's compact Green functions and vortex-sound theory. To further understand the spatial organisation of the acoustically important coherent structures we consider the decomposition of the source term in waves along the cylinder span ($k_2-$modes) by taking the wavenumber transform along the $y_2$-direction. Due to the discontinuity of the sTR-PIV data at the boundaries of the airfoil, the Fourier transform is performed for $-2 \leq y_1 \leq 0$. It reads as, 
\begin{equation}
   \hat  {\hat{S}}\left( y_1, k_2 \right) = \int_{-\infty} ^{\infty} \bar{u}_1\left( y_1, y_2\right)  \hat \omega_3^{\mathrm{2-D}}\left( y_1, y_2, \mathrm{St} \approx 0.38 \right) e^{-\mathrm{i}k_2 y_2} dy_2.
    \label{k2_Transform_source}
\end{equation}

\noindent The parameters of the Fourier transform depend on the spatial discretisation $ \Delta y_2 \approx 0.019 $ with a total of $N_{y_2}=135$ grid points along $y_2$. We perform zero-padding, by centering the signal and using $N_{y_2,\mathrm{fft}} =256$ points. The wavenumber resolution is $\Delta k_2 = \frac{2 \pi }{ N_{y_2} \Delta y_2 } \approx 1.32,$ while according to Nyquist theorem the maximum sampling frequency is $k_{2,\mathrm{max}} = \frac{\pi}{\Delta y_2}$, corresponding to a minimum wavelength $\lambda_2^\mathrm{min}$ (expressed in dimensions) of $$\lambda_2^\mathrm{min} \approx0.19 d.$$
\noindent As a result, the finest length-scales that can be resolved have size comparable to one fifth of cylinder's diameter. Equation \eqref{k2_Transform_source} reads as 

\begin{equation}
   \hat  {\hat{S}}\left( y_1, k_2 \right) = \int_{\mathrm{min}\{y_2\}} ^{\mathrm{max}\{y_2\}} \bar{u}_1\left( y_1, y_2\right)  \hat \omega_3^{\mathrm{2-D}}\left( y_1, y_2, \mathrm{St} \approx 0.38 \right) e^{-\mathrm{i}k_2 y_2} dy_2,
    \label{k2_Transform_source_2}
\end{equation}

\noindent where the integrand of \eqref{k2_Transform_source_2} is complex-valued and we account for positive and negative wavenumbers $k_2$.  We are interested in a description of the wavenumbers that compose the aeroacoustic source. We therefore define the wavenumber spectrum as $\left|   \hat  {\hat{S}}\left( y_1, k_2 \right) \right| ^2$ and is shown in figure \ref{fig:k2_spectra}. A convergence study of the obtained $k_2-$spectra is shown in Appendix \ref{App:k2_spectra}.

It is observed that the $k_2-$spectra of the source term are dominated by four peaks, corresponding to $k_2 \approx \{-10; -5; 0,5\}$ and hence the source term could, in a simplified representation, be described by the associated wavelengths. 

\begin{figure}
\centering
   \includegraphics[width=0.7\textwidth]{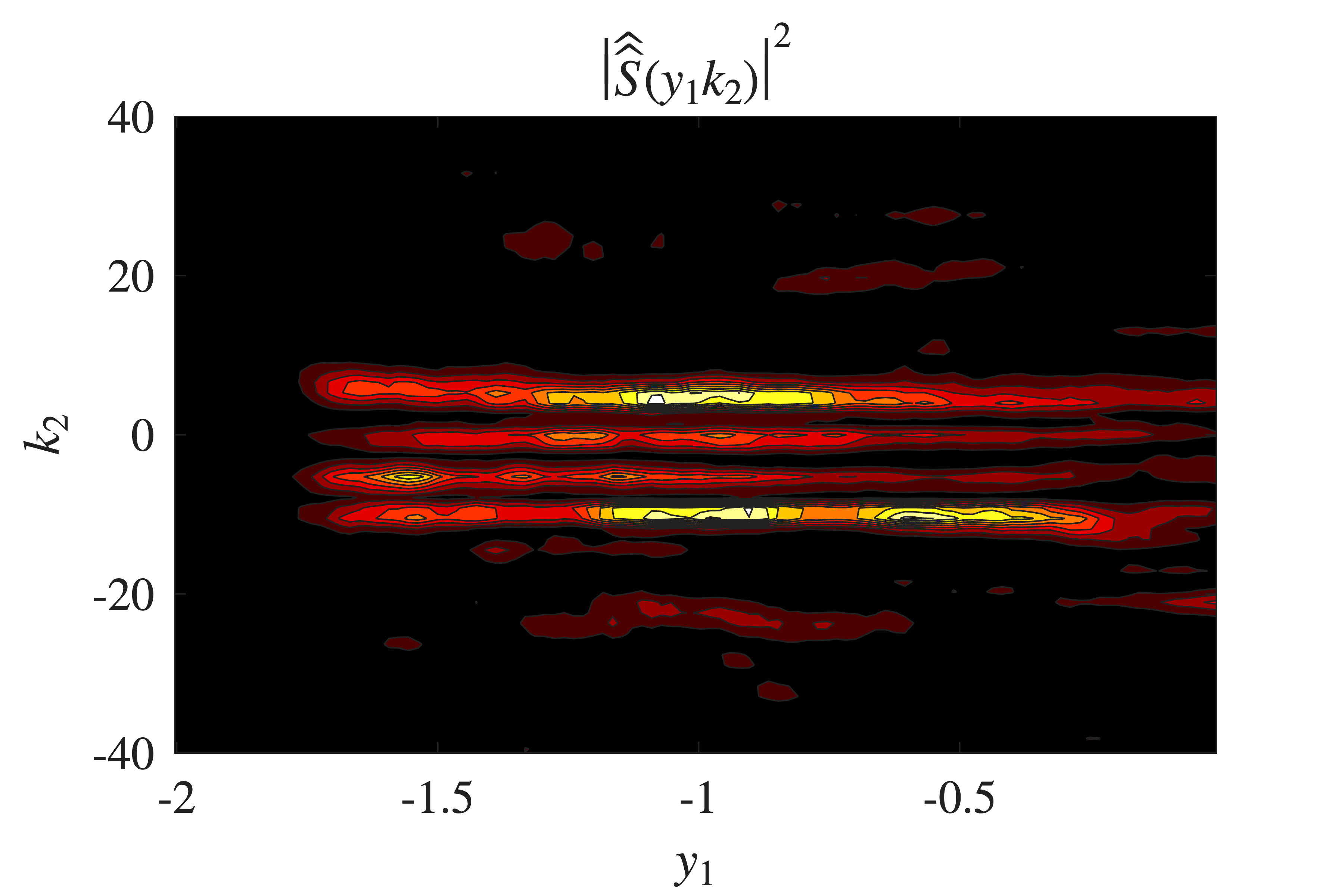}
    \caption{Wavenumber spectrum $\left|   \hat  {\hat{S}}\left( y_1, k_2 \right) \right| ^2$ for $\mathrm{St}=0.38$.}
\label{fig:k2_spectra}
\end{figure}

\subsection{Semi-empirical source model}
\label{subsec:semi_empirical}
The difference between $k_2-$modes that compose the source's amplitude (energy) and those that contribute to the acoustic field, is that the latter get propagated by the Green function and the integration over the source region. In what follows, we develop a semi-empirical source model based on the $k_2$-decomposition to identify the combination of $k_2-$modes that adequately capture the measured acoustic field via the proposed aeroacoustic formulation of eq. \eqref{ff_pressure2}. The procedure we follow is divided in the following steps: 

\subsubsection*{Step 1: Streamwise averaging}
To get an overall representation of the source term, the wavenumber-spectrum $\left|   \hat  {\hat{S}}\left( y_1, k_2 \right) \right| ^2$ is averaged over all-streamwise positions for $-2 \leq y_1 \leq0$

\begin{equation}
	\hat A(k_2) =\frac{ \int_{-2}^{0} \left|   \hat  {\hat{S}}\left( y_1, k_2 \right) \right| ^2 \mathrm{d}y_1}{\int_{-2}^{0}\mathrm{d}y_1}.
\label{eq:averaged_k2}
\end{equation}

\begin{figure}
    \centering
    \begin{minipage}{0.48\textwidth}
        \centering
               \includegraphics[width=\linewidth]{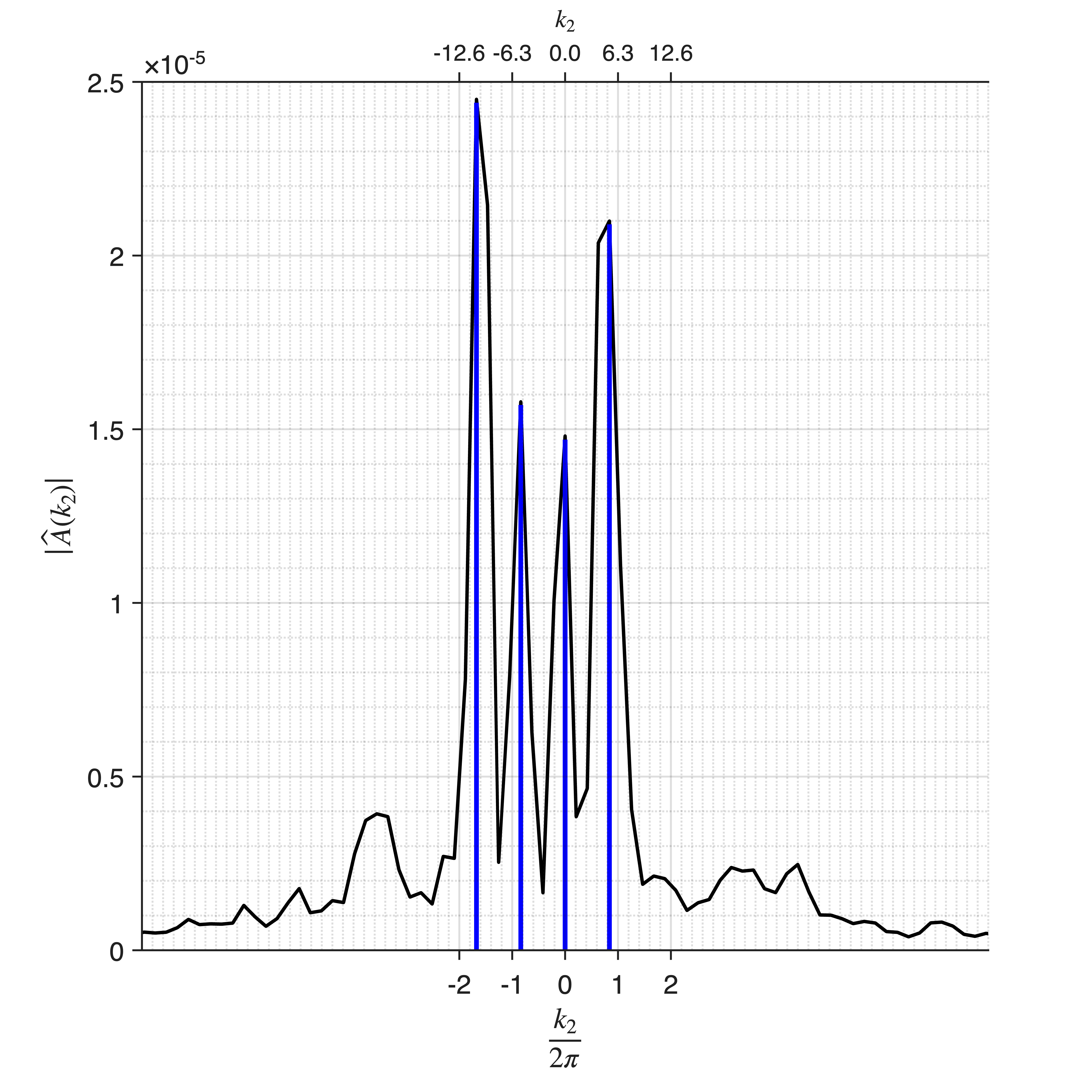}
  \caption{Streamwise-averaged $k_2-$spectrum (black) and dominant wavenumbers (blue).}
\label{fig:k2_spectra_fit}   
    \end{minipage}
    \hfill
    \begin{minipage}{0.48\textwidth}
        \centering
      \includegraphics[width=\linewidth]{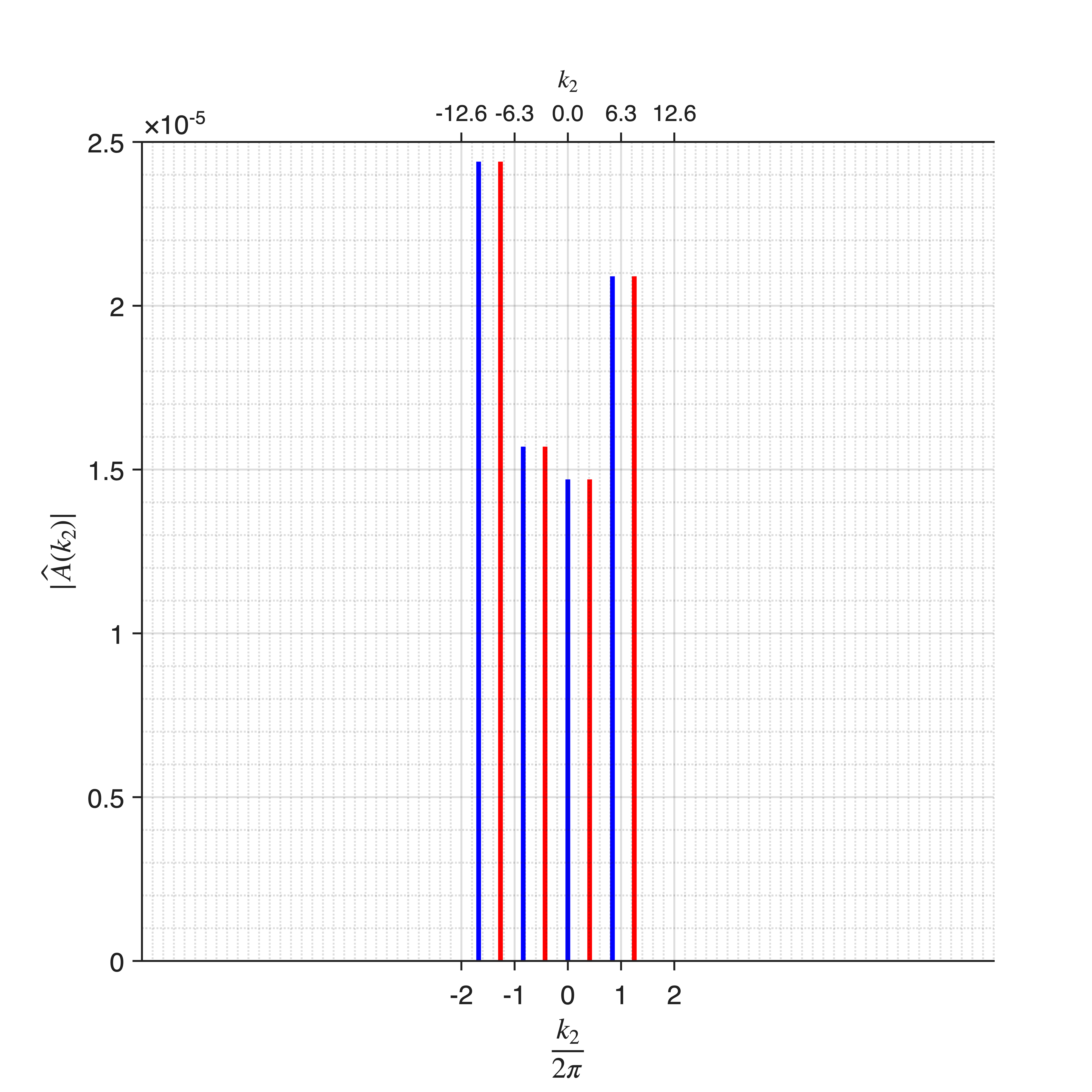}
        \caption{Modelled (red) vs experimentally-obtained (blue) wavenumber spectra.}
\label{fig:k2_spectra_ideal}
    \end{minipage}
\end{figure}

It is noted that the wavenumber spectrum shows a slight asymmetry around $k_2=0.$ The latter is an artefact of the experimental set-up associated with an inclination of the main vortices of the cylinder's wake as shown in Movie 1 of the instantaneous  velocity field provided in the supplementary files or in the dominant spectral-proper-orthogonal-decomposition modes of figure 15 of \cite{do_amaral_perpendicular_2025}. This artefact is observed also in other experimental campaigns conducted in the same wind tunnel \citep{prinja2025experimentally}.
\subsubsection*{Step 2: Source representation as a Fourier-series}
Following the previous analysis we can now construct a semi-empirical source model based on the dominant wavenumbers shown in figure \ref{fig:k2_spectra_fit}. In what follows, we consider the ideal case of a symmetric wavenumber spectrum, where the inclination of the vortex rolls is neglected. The latter assumption is valid since a variation of the relative angle of the vortex and the leading edge is not expected to create noise as shown in figure \ref{Interaction_verticalhorizontal}, where the acoustic spectra of the airfoil with and without angle of incidence do not show significant differences. Figure \ref{fig:k2_spectra_ideal} depicts the simplified modelled (symmetric) and experimentally-obtained (asymmetric) wavenumber-spectra. 


The analysis is focused at $\mathrm{St}=0.38$, hence the source-model $\hat{\mathcal{S}}$ depends only on the source-coordinates $y_1,y_2,y_3$. The airfoil-spanwise dependence of the source was analysed in \S \ref{subSection:Datamodel}, while the streamwise dependence is driven by the non-dimensional convective wavenumber $$k_1 = k/M =  2 \pi \mathrm{St}\frac{c}{d}. $$ \noindent It then follows, 
\begin{equation}
\hat{\mathcal{S}}(\vec{y}) = 
\sum_{m=-128}^{128} 
\Bigg\{ 
\sqrt{\hat{A}(m \Delta k_2)} \, 
e^{\mathrm{i} m \Delta k_2 y_2} 
\Bigg \}
\left( 
e^{\mathrm{i} k_1 y_1} \, 
e^{-\frac{y_3^2}{L_3^2}} 
\right),
\end{equation}
\noindent where the summation over $m$ corresponds to the summation over wavenumbers, obtained by the Fourier transform in $y_2$, such that $k_2=m \Delta k_2$. $L_3$ is the support of the vorticity fluctuation field along the airfoil-span (\S \ref{subSection:Datamodel}). However, as shown in figure \ref{fig:k2_spectra_fit}, the wavenumber spectrum is dominated by four $k_2-$modes and hence can be approximated as
\begin{equation}
\hat{\mathcal{S}}(\vec{y}) \approx \sum_{k_2=\{-7.6,-2.6,2.6,7.6\}}  \sqrt{\hat{A}(k_2)}  e^{\mathrm{i}  k_2 y_2}  \left(  e^{\mathrm{i} k_1 y_1} e^{-\frac{y_3^2}{L_3^2}} \right).
\label{eq:source_model}
\end{equation}

\noindent

\subsubsection*{Step 3: Acoustic based identification of the aeroacoustic source}
\begin{figure}
\centering
   \includegraphics[width=0.95\textwidth]{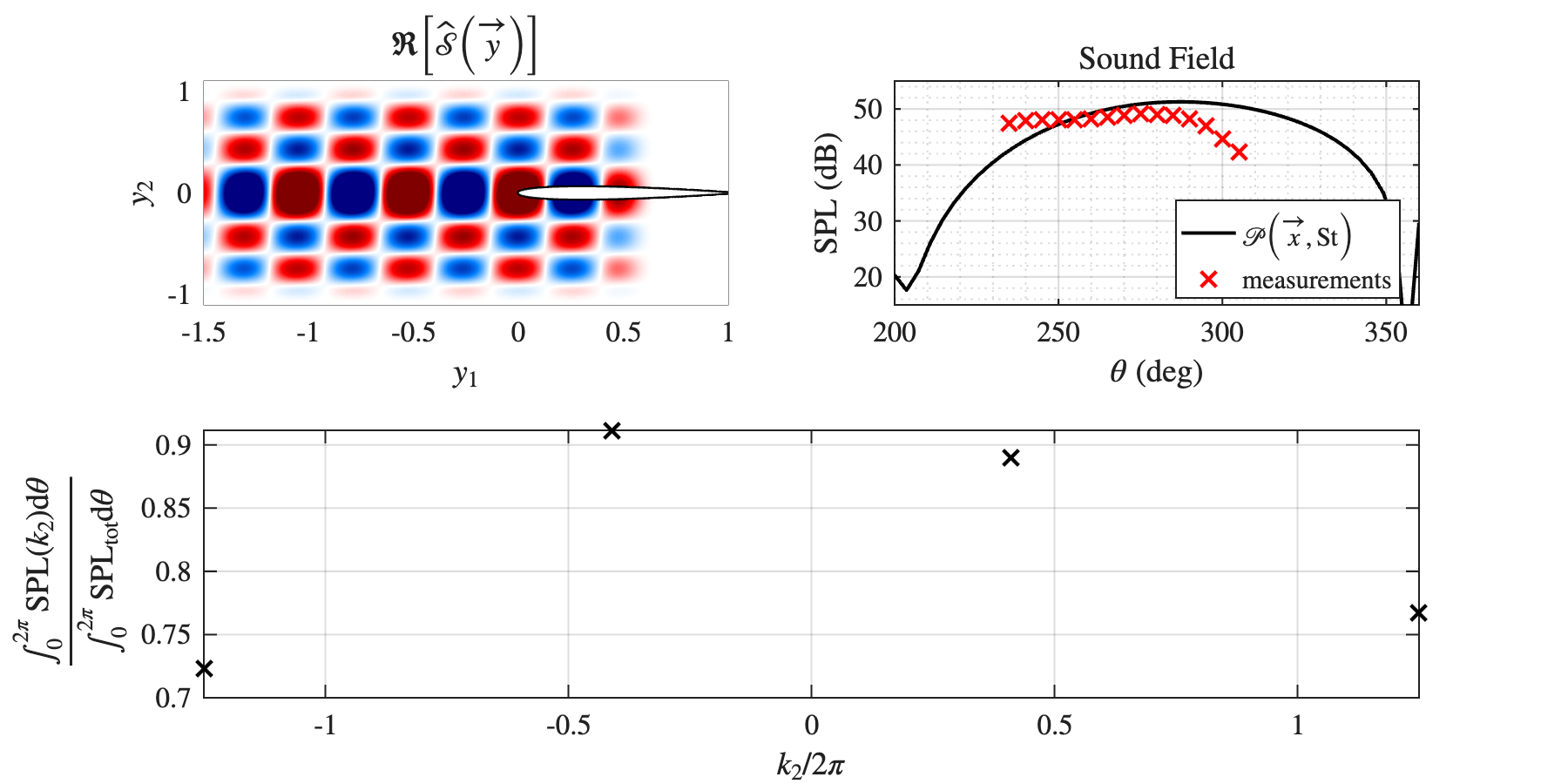}
    \caption{Semi-empirical source term (left) and acoustic field (right), computed by \eqref{ff_pressure3} vs measured SPL. The bottom diagram corresponds to the overall contribution of each wavenumber in the total sound pressure level (eq. \eqref{eq:mu}).}
\label{fig:Error_comb}
\end{figure}

Combining equations \eqref{ff_pressure2}, \eqref{eq:source_model} we propose the following data-informed, semi-empirical aeroacoustic model. The wavenumber analysis is restricted up to the leading edge of the airfoil and therefore, we choose the parameters  $y_{1,c}=-0.5, L_1=1$ of the window-function \eqref{eq:super_Gaussian}  to consider the source-term in the vicinity of the leading edge. The integration is performed over a symmetric domain $D:  \{ -4 \leq y_2 \leq 4, -4 \leq y_1 \leq 4 \}$
\begin{equation}
\begin{aligned}
      \mathcal{P} \left(\vec{x},\mathrm{St}\right) = \frac{i  \rho  U_\infty^2 e^{\mathrm{i}k|\vec{x}|}  }{4 \pi  |\vec{x}|}   \frac{x_2}{|\vec{x}|}  \frac{ \mathcal{L}_3(x_3,\mathrm{St})}{d}  &  \sum_{k_2} \Biggl\{    \iint_{D}    \sqrt{\hat{A}(k_2)}  e^{\mathrm{i}  k_2 y_2}   e^{\mathrm{i} k_1 y_1}  \mathcal{W}(y_1,y_2)\\
     &\times \left( \frac{x_1}{|\vec{x}|} \frac{\partial Y_1(y_1,y_2)}{\partial y_2}+ \frac{x_2}{|\vec{x}|} \frac{\partial Y_2(y_1,y_2)}{\partial y_2} \right) e^{-\mathrm{i}k \left(\frac{x_1 Y_1}{|\vec x|}+\frac{x_2 Y_2}{|\vec{x}|}\right)} \mathrm{d}y_1 \mathrm{d}y_2 \Biggr\},
\end{aligned}
\label{ff_pressure3}
\end{equation}

\noindent where $k_2 =\{-7.6,-2.6,2.6,7.6 \}.$ The top plot of Figure \ref{fig:Error_comb} shows the modelled, semi-empirical source (left) and the computed acoustic field of \eqref{ff_pressure3}, using all dominant wavenumbers compared to the acoustic measurements (right). The modelled acoustic field lies overall in acceptable agreement with the experimental data. In the bottom plot of Figure \ref{fig:Error_comb}, the sound levels are computed for each wavenumber and an integration over the angular location is performed.  To evaluate the contribution of each wavenumber in the acoustic field, we consider  the metric
\begin{equation}
\mu = \frac{\int_{0}^{2 \pi}\mathrm{SPL}(k_2) \mathrm{d}\theta}{\int_{0}^{2 \pi}\mathrm{SPL}_{\mathrm{tot}} \mathrm{d}\theta},
\label{eq:mu}
\end{equation}
 
\noindent where $\mathrm{SPL}_{\mathrm{tot}}$ corresponds to the sound pressure level computed form the model of eq. \eqref{ff_pressure3}. It is thus shown that the acoustically important wavelengths are comparable to the size of the chord, $\lambda_2 \approx 0.80c; 2.45c$ that corresponds to $\approx 4d;12d$ respectively. 

The data-informed, simplified, semi-empirical source model for the Lamb-vector shows similar length-scales ($\approx c$) with the acoustically important $u_2$-unsteady-velocity coherent structures, extracted using extended spectral proper orthogonal decomposition (ESPOD), presented in the companion study of \citet[figure 13]{do_amaral_perpendicular_2025}. Notice that only a qualitative comparison of these studies is allowed since the one relies on semi-analytical, linearised, data-informed modelling and the other on post-processing techniques operating on experimental data. However, both result in the same conclusion that coherent structures related to the unsteady upwash/downwash velocity fluctuations of length scales comparable to the chord's size drive the interaction noise. 

 In the bench-marked rod-airfoil problem \citep{jacob2005rod}, where the spans of the cylinder and the airfoil are aligned, the von Kármán vortex shedding is strongly associated with the acoustic field, whereas our results suggest that higher-order cylinder-span Fourier modes with airfoil-span-aligned vorticity are important. Based on the observations and analysis presented previously, the latter, acoustically-important flow structures appear at a characteristic frequency $St\approx 0.38$, i.e. cylinder's drag force frequency. 

Cylinder wakes have been extensively studied in the literature and various coherent structures linked to flow instabilities have been identified. A detailed review and study of those three-dimensional instabilities can be found in the work of \citet{williamson1996mode}. We hypothesise that the observed, acoustically-important structures may arise due to secondary instabilities of the two-dimensional von Kármán structures. \citet{williamson1988existence} and  \citet{barkley1996three} have demonstrated that there exist two types of flow instabilities evolving in the streamwise direction of the flow which lead to a three-dimensional wake transition, called mode A and mode B.  Mode A is an instability of the vortex core associated with the primary von Kármán vortices. The streamwise vortices due to mode A create vortex loops of typical wavelength corresponding to $\approx 5d$ that interact with the primary vortices, leading to deformations of the latter \citep{williamson1996mode}.

Those instabilities have been studied at very low Reynolds numbers and cannot be directly applied in the present problem. However, given the success of linear mean-flow stability analyses (cf. literature on turbulent jets \citep{jordan2013wave,cavalieri2019wave}) it can be conjectured that a mechanism similar to that of the mode A may be active in the turbulent cylinder wake and, supported by our analysis, important for sound generation. It is thus hypothesised that the sound is generated due to secondary instabilities in the wake of the cylinder. These wake structures are linked to the three-dimensional nature of the flow field, and as observed by the experimental investigation show a peak at $\mathrm{St}=0.38$ \citep{do_amaral_perpendicular_2025}. 
 
\section{Conclusions}
Experiments took place in order to examine the sound generated when the wake of a cylinder interacts with a downstream airfoil when the two objects have their spans orthogonally aligned. The present set-up differs substantially to classical studies related to cylinder noise or cylinder-wake/airfoil aeroacoustics because of the three dimensional nature of the cylinder's wake and in consequence, due to the fact that the mechanisms of sound generation are associated with the unsteady cylinder-span-aligned velocity fluctuations acting on the leading edge.

Based on the experimental results, a physical analysis of the noise generation mechanisms was proposed by deriving a semi-analytical expression based on the theory of vortex sound and a linearisation of the source term. The derived model assumes that the airfoil is acoustically compact and Howe's compact-Green-function theory is employed to account for the acoustic scattering. The source-term is obtained by the experimental data and an estimation of the acoustic field with acceptable accuracy is obtained via the proposed model. 

Further analysis was carried out to explore the physics of the sound generation mechanisms, focusing on the role of the airfoil-span-aligned vorticity in the acoustic field. The aeroacoustic source was decomposed in spatial waves extending along the span of the cylinder and a semi-empirical model based on the experimental data was constructed to evaluate the contribution of each wave to the sound field. The data-informed, semi-empirical model showed that flow structures with a size comparable to the chord of the airfoil drive the interaction noise, which lies in agreement with similar observations reported in the work of \citet{do_amaral_perpendicular_2025}. As a result, the topological properties of the vorticity field that cause the generation of sound waves have been identified and their link to secondary cylinder wake instabilities was conjectured. A detailed analysis of the secondary instabilities of the cylinder's wake would shed further light in the true origin of these acoustically important cylinder's wake structures.

\backsection[Supplementary data]{\label{SupMat}Movies of the fluctuation field in both planes (Movie1, Movie 2) are available at ...}

\backsection[Acknowledgements]{The authors are grateful to the technical staff at Institute Pprime for their invaluable support. More specifically we would like to thank Laurent Philippon for his contributions in the set-up of the measurement systems (sTR-PIV and microphones) and the operation of the wind tunnel, Pascal Biais for manufacturing the elements of the experimental configuration and Damien Eysseric for his contributions in the set-up of the sTR-PIV instrumentation.}

\backsection[Funding]{This work was supported by the DGAC (Direction Générale de l'Aviation Civile), by the PNRR (Plan National de Relance et de Résilience Français) and by NextGeneration EU via the project  MAMBO (Méthodes Avancées pour la Modélisation du Bruit moteur et aviOn).}

\backsection[Declaration of interests]{ The authors report no conflict of interest.}


\backsection[Author ORCIDs]{  M.I. Spiropoulos, https://orcid.org/0009-0004-6462-9585; F.R. Amaral, https://orcid.org/0000-0003-1158-3216 ; P. Jordan, https://orcid.org/0000-0001-8576-5587}

\newpage

\appendix

\section{Appendix: Details on the sPIV-experimental set-up}
\label{App_PIV}

In the present Appendix further technical details regarding the PIV set-up are mentioned.
\subsection{Instrumentation}
The PIV-set-up is composed of:
\begin{itemize}
    \item Camera-system
        \begin{itemize}
                    \item Two cameras: \textit{Vision Research Phantom $v2640$}
                    \begin{itemize}
                        \item Resolution: $1024 \times 1024$ pixels
                        \item Resolution speed: $3000$ images$/$s
                    \end{itemize}
                     \item Two lenses: \textit{Nikon AF Micro-Nikon 60} mm $1:2.8 $D 
        \end{itemize}
    \item Laser-system 
        \begin{itemize}
     
            \item  \textit{Continuum MESA PIV}
            \begin{itemize}
                \item Wavelength: $532$nm
                \item Nominal power: $2 \times 9$ mJ at $152$ $\mu$s
                \item Lens' diameter:  $5$ mm
                \item Pulse duration: $<150$ ns
                \item Frequency range: $1 \leq f_\text{MESA} \leq 40$ kHz
            \end{itemize}
            \item \textit{Photonics Industries DMX 150-532 DH}
            \begin{itemize}
                 \item Wavelength: $532$nm
                \item Nominal power: $2 \times 15$ mJ at $152$ $\mu$s
                \item Lens' diameter:  $4.5$ mm
                \item Pulse duration: $200$ ns
                \item Frequency range: $1 \leq f_\text{DMX} \leq 50$ kHz
            \end{itemize}
       \end{itemize}
         
\end{itemize}
\subsection{Configuration of the lasers for both planes}

\begin{figure}
\centering
   \includegraphics[width=0.79\textwidth]{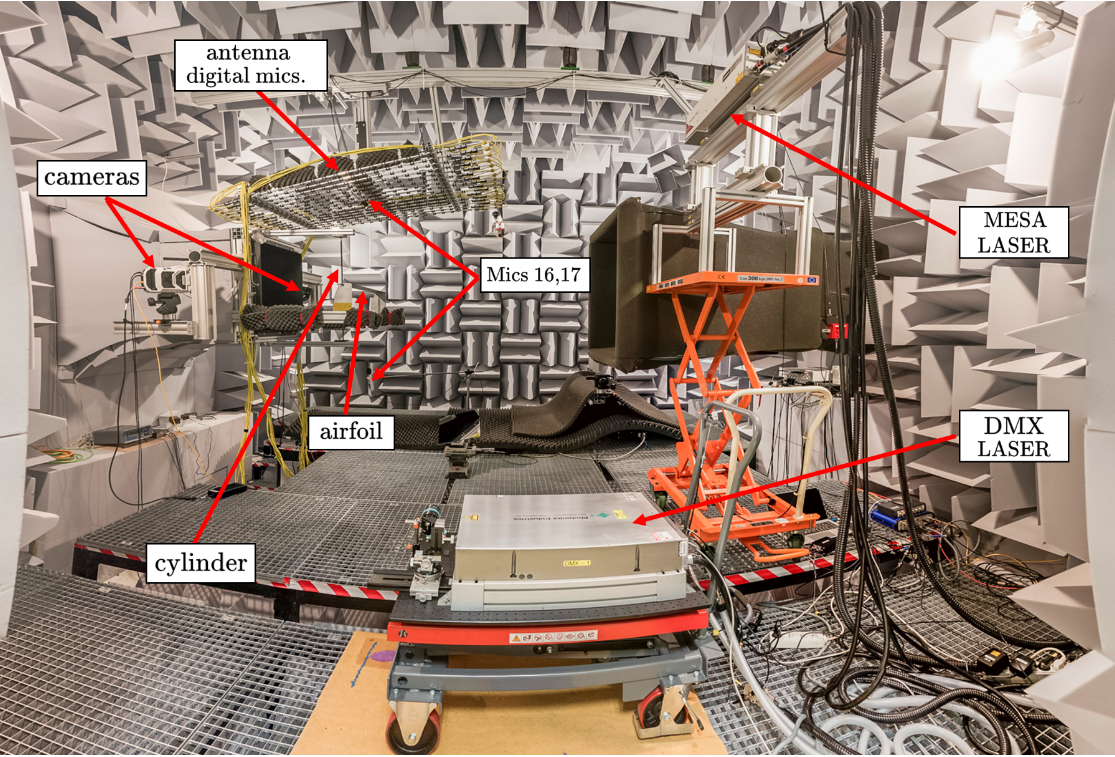}
    \caption{Experimental set-up of the s-PIV campaign for the measurements taken on the $y_1-y_2$ plane.}
\label{SETUP_Y1Y2}
\end{figure}

\begin{figure}
\centering
   \includegraphics[width=0.79\textwidth]{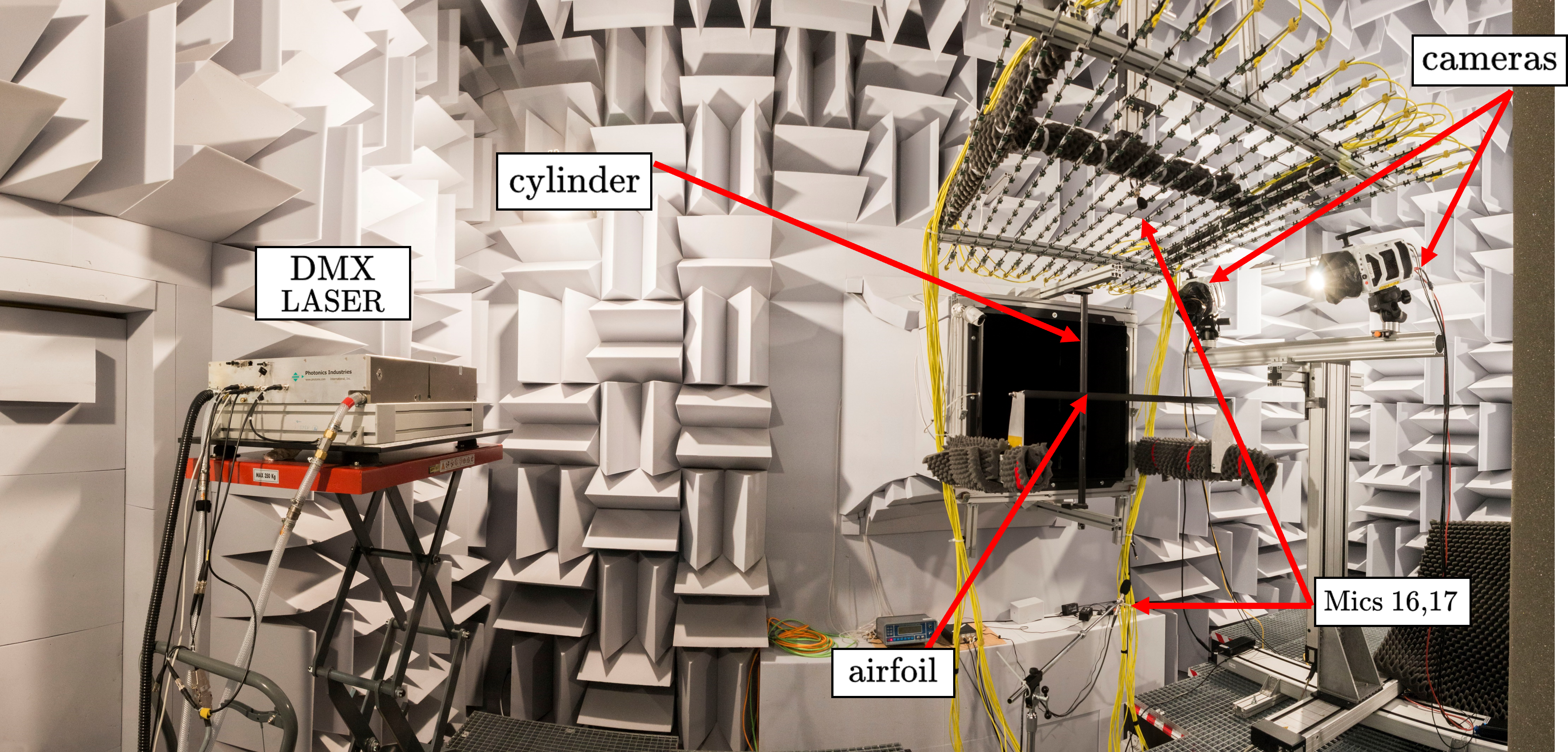}
    \caption{Experimental set-up of the s-PIV campaign for the measurements taken on the $y_1-y_3$ plane.}
\label{SETUP_Y1Y3}
\end{figure}

Both Laser-systems were used for the $y_1-y_2$ plane, while only the DMX laser was used for the $y_1-y_3$ plane. The laser beam was directed by mirrors in the experimental setup concerning the measurements on the $y_1-y_2$ plane. The positioning of the PIV system for both planes is shown on Figs. \ref{SETUP_Y1Y2}, \ref{SETUP_Y1Y3} for the $y_1-y_2$, $y_1-y_3$ planes respectively. 

\begin{figure}
\centering
   \includegraphics[width=0.89\textwidth]{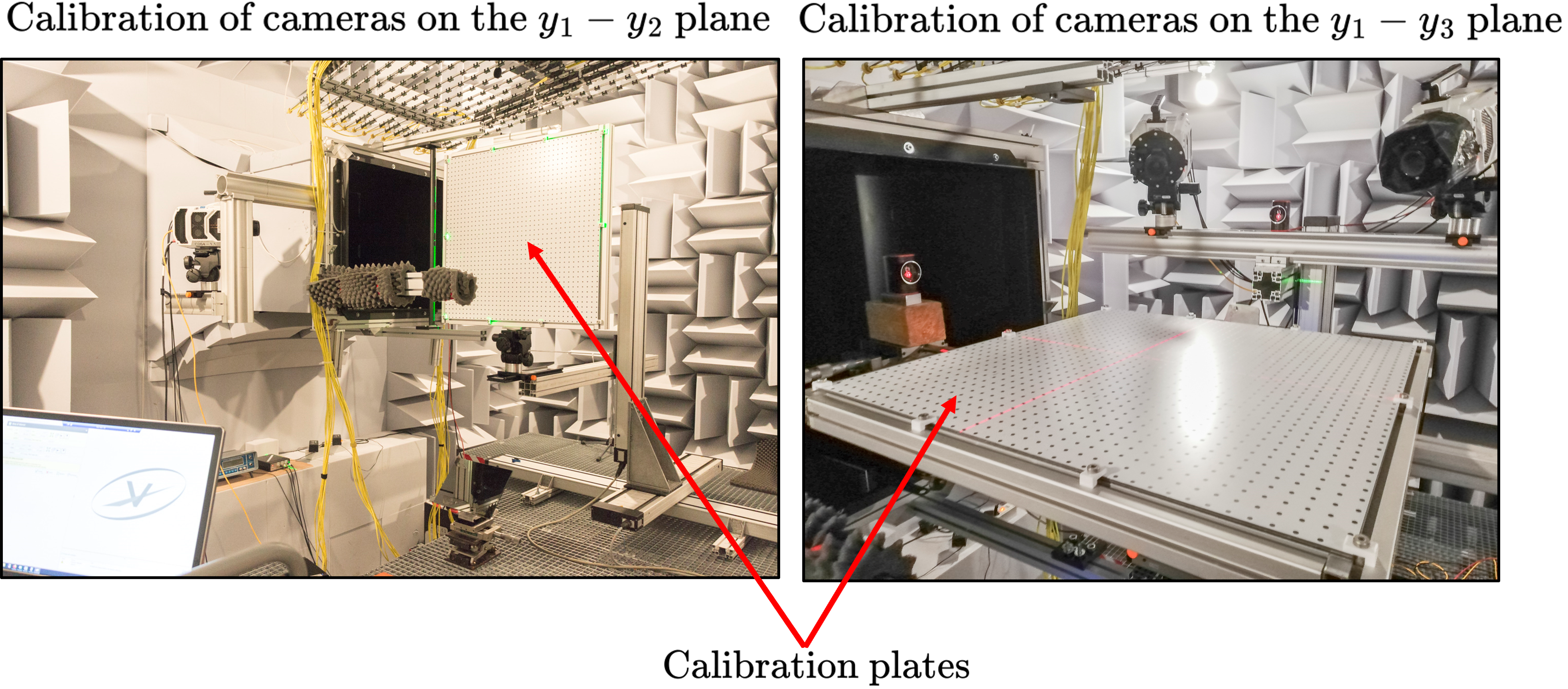}
    \caption{Calibration of cameras by using the calibration plates for the $y_1-y_2$ plane (left) and the $y_1-y_3$ plane (right).}
\label{CallibrationPlates}
\end{figure}

The calibration of the cameras was done by using two plates of $750 \times 750$ mm$^2$ as shown on figure \ref{CallibrationPlates}. The polynomial model is used with a scaling factor of 4.29 pixel/mm and the Pinhole model with a scaling factor 5,05 pixel/mm for the $y_1-y_2$ and $y_1-y_3$ planes respectively.

\newpage
\subsection{Seeding of the particles}
\label{App:Shadow_zone}
\begin{figure}
\centering
   \includegraphics[width=0.99\textwidth]{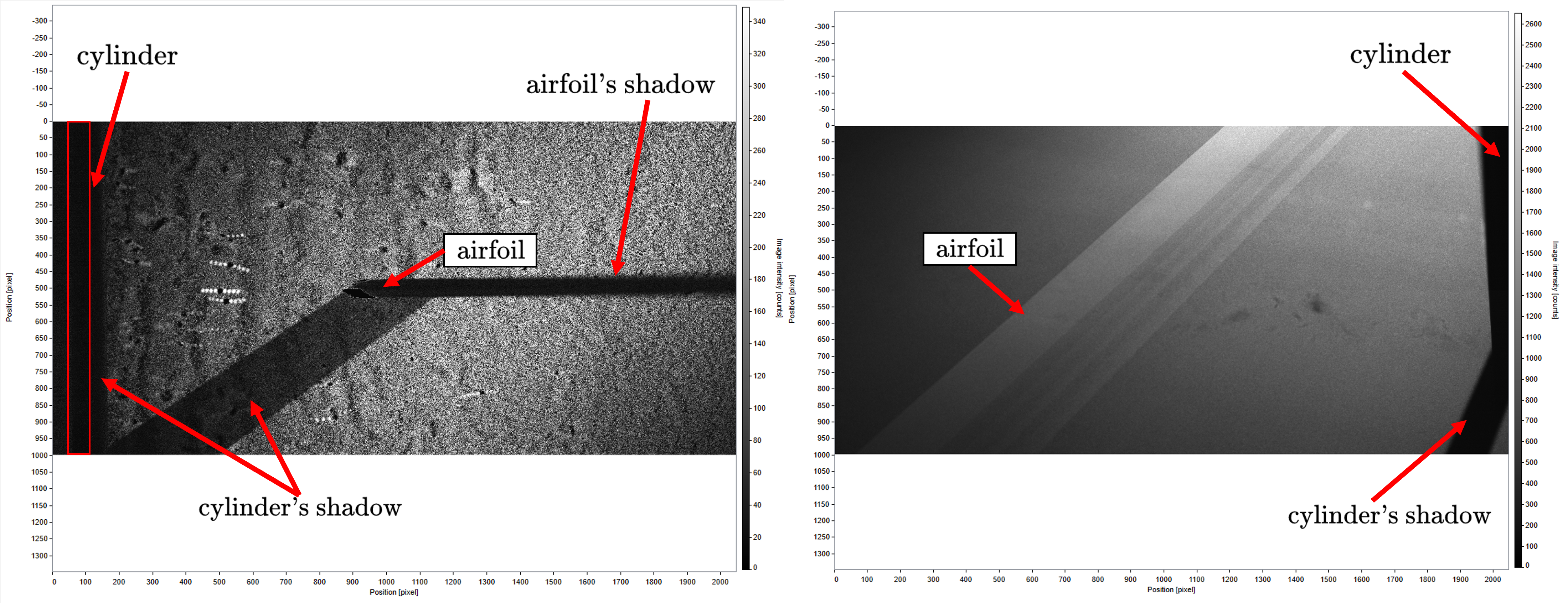}
    \caption{Raw sTR-PIV-images as obtained from one of the two cameras for the $y_1-y_2$ plane (left) and the $y_1-y_3$ plane (right).}
\label{Seeding_y1y3_y1y3}
\end{figure}

A fog-machine (Antari Z3000) was used to seed the air particles with heavy smoke (ALGAM LIGHTING). A random snapshot of the image as obtained by the cameras is shown on figure \ref{Seeding_y1y3_y1y3}. The luminosity of the image is represented in grey scale, with white corresponding to maximum luminosity and dark to a poor quality of the image. It can be seen, that in the vertical plane (figure \ref{Seeding_y1y3_y1y3} left), the shadow of the cylinder influences the quality of the image and therefore the results in this region are sensitive to errors, e.g. when computing the time-averaged and root-mean-square values of the velocity and vorticity fields (see Figs \ref{u_rms}, \ref{Mean_vorticity}, \ref{vorticity_rms}).

\subsection{Image processing}
The sampling of the images was done at $6$ and $6.25$ kHz for the $y_1-y_2$ and $y_1-y_3$ planes respectively. The time between two laser-pulses depends on the flow-speed, for the case-study presented here, a $\Delta t=40$ $\mu$s was chosen. The lasers were set at their maximum power output for both pulses.

After the sampling, the following procedure for the image processing using the DAVIS software was followed. At first, it was verified that the lasers and the calibration plates are aligned by using the self-calibration toolbox. Furthermore, a sliding-minimum window filter of 9 images was applied. To determine the vector fields, the multi-pass PIV calculation is employed as follows i) 2 passes with a square-shaped window with area of $64 \times 64$ pixels$^2$ and ii) with 2 passes of automatic shape with an area of $16 \times 16$ pixels$^2$. The overlap is set at 50$\%$. Finally, the vector-fields showing a correlation less than $0.3$ were removed and interpolation between the validated vectors was used to extract the vector-field.

\section{Simplification of the source term based on the experimental data}
\label{Lamb}
As demonstrated in \S \ref{subSec:TVS}, the acoustic dipole, oriented parallel to the cylinder's span, is linked to the interaction noise and can be expressed as 

\begin{equation}
   \begin{aligned}
  \hat p= \rho \int_V \left[\left( \bar{\omega}_1  \hat  u_3 + \hat  \omega_1 \hat  u_3\right)   -  \left(\bar{\omega}_3 \bar{u}_1  +\hat   \omega_3  \bar{u}_1+\bar{\omega}_3 \hat  u_1 + \hat  \omega_3 \hat u_1\right) \right]  \frac{\partial G}{\partial y_2} \mathrm{d}V.
    \end{aligned}
     \label{App:Dipoles3}
\end{equation}

In what follows, we compare the order of magnitude of the fluctuating quantities that comprise the source term. To do so we assume that the magnitude frequency-domain fluctuation quantities ($\hat q_i$), $q$ being an arbitrary quantity (vorticity/velocity), can be estimated by their root-mean-square values $\left( q^{\mathrm{rms}}\right)$. Hence for the subsequent analysis, we replace
\begin{align*}
\hat \omega_i  &\quad \text{with} \quad \frac{\omega^{\text{rms}}_id }{U_\infty},\\
\hat u_i  & \quad \text{with} \quad \frac{u^{\text{rms}}_i }{U_\infty},\\
\end{align*}
\noindent where the rms and time-averaged quantities are kept dimensional. Notice that this notation is only kept in this Appendix, while the equations related to the far-field acoustic pressure, as \eqref{App:Dipoles3}, will keep the same notation as in the main text.

\begin{figure}
    \centering
    
    \begin{subfigure}{\textwidth}
        \centering
        
        \begin{subfigure}{\textwidth}
            \centering
            \includegraphics[width=\textwidth]{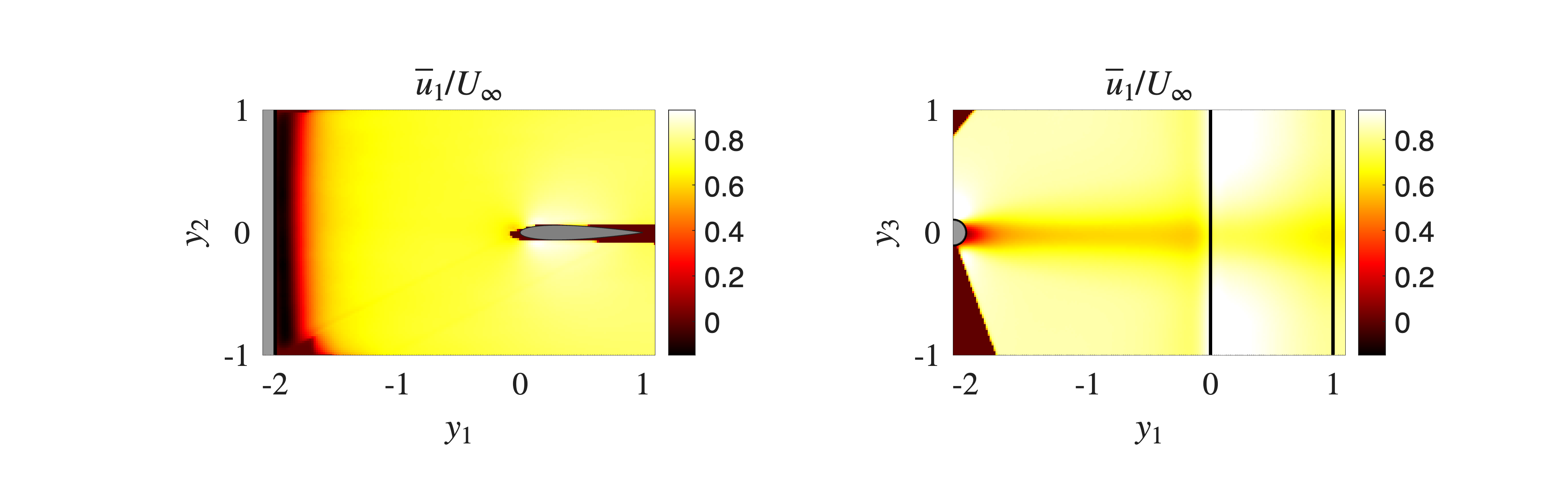}
        \end{subfigure}

        \begin{subfigure}{\textwidth}
            \centering
            \includegraphics[width=\textwidth]{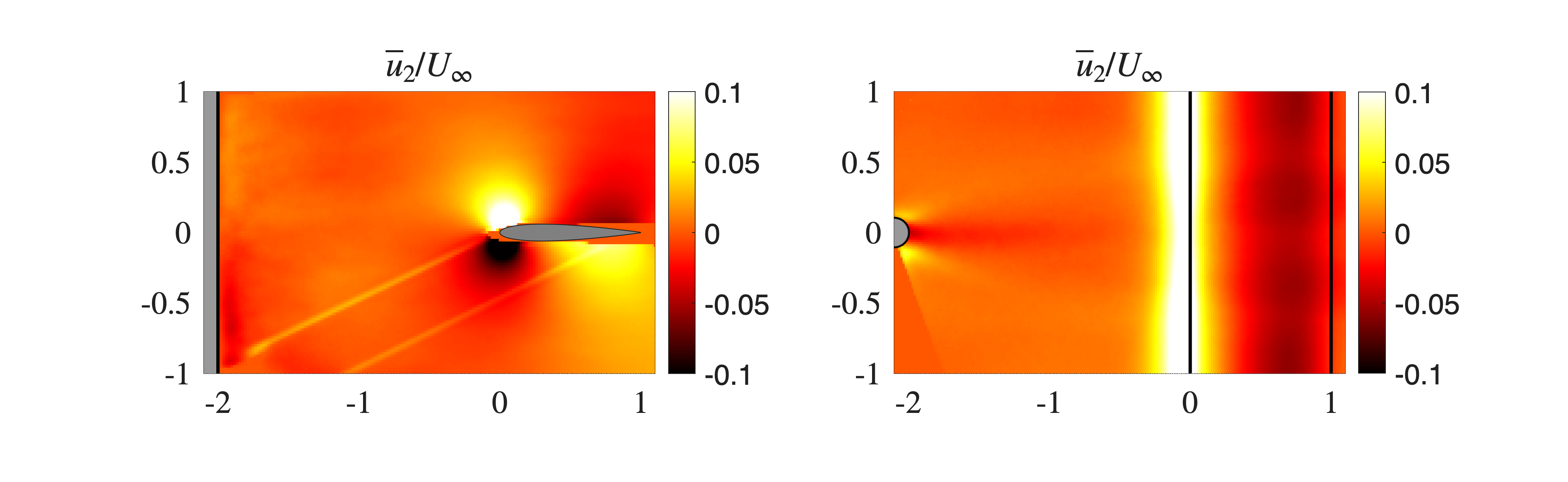}
        \end{subfigure}

        \begin{subfigure}{\textwidth}
            \centering
            \includegraphics[width=\textwidth]{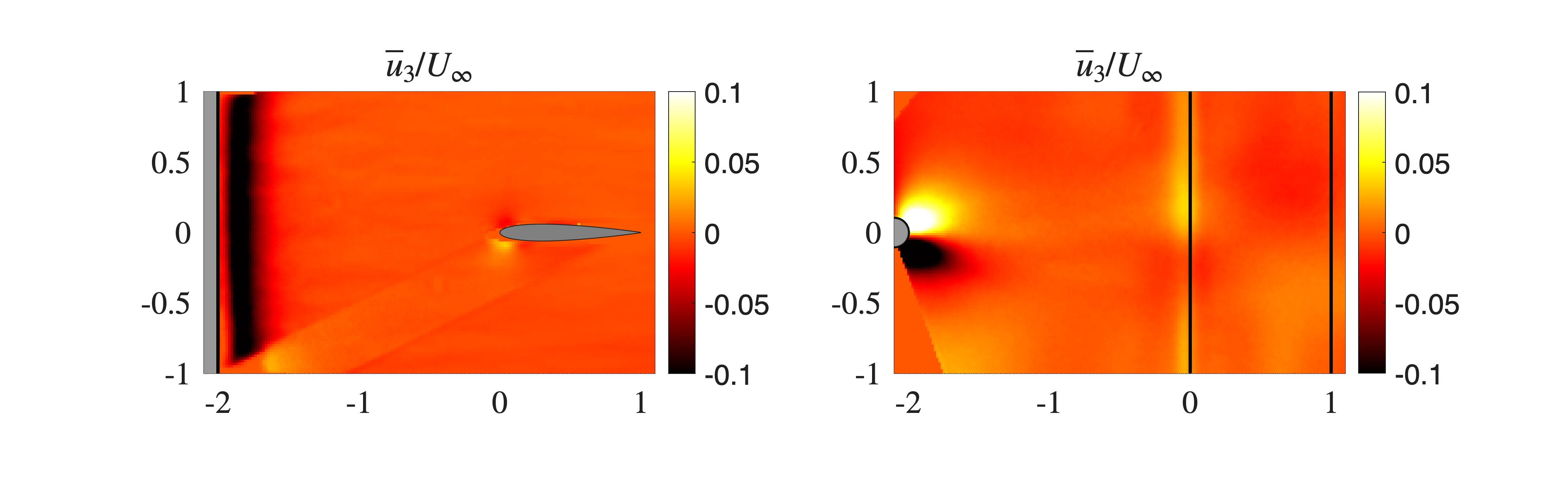}
        \end{subfigure}
  \end{subfigure}
    
 \caption{Mean velocity field components, as obtained from the TR-PIV measurements in the $y_1-y_2$ (left column) and the $y_1-y_3$ (right column) plane.  The top, middle and bottom rows correspond to the streamwise $\left( \bar{u}_1 \right)$, cylinder-span aligned $\left( \bar u_2 \right) $ and airfoil-span algned $\left( \bar u_3 \right) $ mean velocity components.}
\label{meanV} 
  \end{figure}

The streamwise, mean velocity component is of an order of magnitude more significant than the corresponding cylinder-span and airfoil span-aligned mean velocity components. This can be shown in figure \ref{meanV}, where the 3-component, mean velocity field is presented in both planes.


To provide a reasoning for the following approximations we introduce the following notation for the mean velocity components 
\begin{equation}
\begin{aligned}
\frac{\bar u_1}{U_\infty} & \approx o(1) \\
\frac{\bar u_3}{U_\infty} &  \approx o(\epsilon),
\end{aligned}
\label{Scaling_mean}
\end{equation}

\noindent where $\epsilon \approx 0.1$ is chosen based on the measurements, see for instance figure \ref{meanV}. Similarly based on the mean vorticity fields of figure \ref{Mean_vorticity} it is observed that the mean airfoil-span-aligned vorticity $\bar \omega_3 d/U_\infty  \approx o(\epsilon)$.

 For the velocity-fluctuation field (figure \ref{u_rms}) we observe that near the interaction region the root-mean-square velocity scales as
\begin{equation}
\begin{aligned}
\frac{u_1^\text{rms}}{U_\infty} &\approx o(\epsilon) \\
\frac{ u_2^\text{rms}}{U_\infty} &\approx o(\epsilon)\\
\frac{ u_3^\text{rms}}{U_\infty} &\approx o(\epsilon),
\end{aligned}
\label{Scaling_rms}
\end{equation}
\noindent while figure \ref{vorticity_rms} shows that the root-mean-square of the airfoil-span-aligned vorticity near the interaction region scales as

\begin{equation}
\frac{\omega_3^\text{rms}d}{U_\infty}  \approx o(1),
\label{Scaling_rms_v}
\end{equation}
 \noindent It then follows, 
 
 \begin{equation}
\begin{aligned}
\bar{\omega}_1 u_3^{\mathrm{rms}}\frac{ d}{U_{\infty}^2}  &\approx o(\epsilon) \frac{\bar{\omega}_1 d}{U_{\infty}}\\
\omega^\text{rms}_1  u^\text{rms}_3\frac{ d}{U_{\infty}^2} &\approx o(\epsilon) \frac{\omega^\text{rms}_1 d}{U_{\infty}} \\
\bar{\omega}_3  \bar{u}_1\frac{ d }{U_\infty^2}  &\approx o(\epsilon)\\
\omega^\text{rms}_3\bar{u}_1 \frac{ d}{U_{\infty}^2} &\approx o(1) \\
\bar{\omega}_3 u^\text{rms}_1\frac{ d}{U_{\infty}^2}   &\approx o(\epsilon^2) \\
\omega^\text{rms}_3 u_1^{\mathrm{rms}} \frac{ d}{U_\infty^2} &\approx o(\epsilon) \\
\end{aligned}
\label{Scaling_rms}
\end{equation}

\begin{figure}
\centering
   \includegraphics[width=0.85\textwidth]{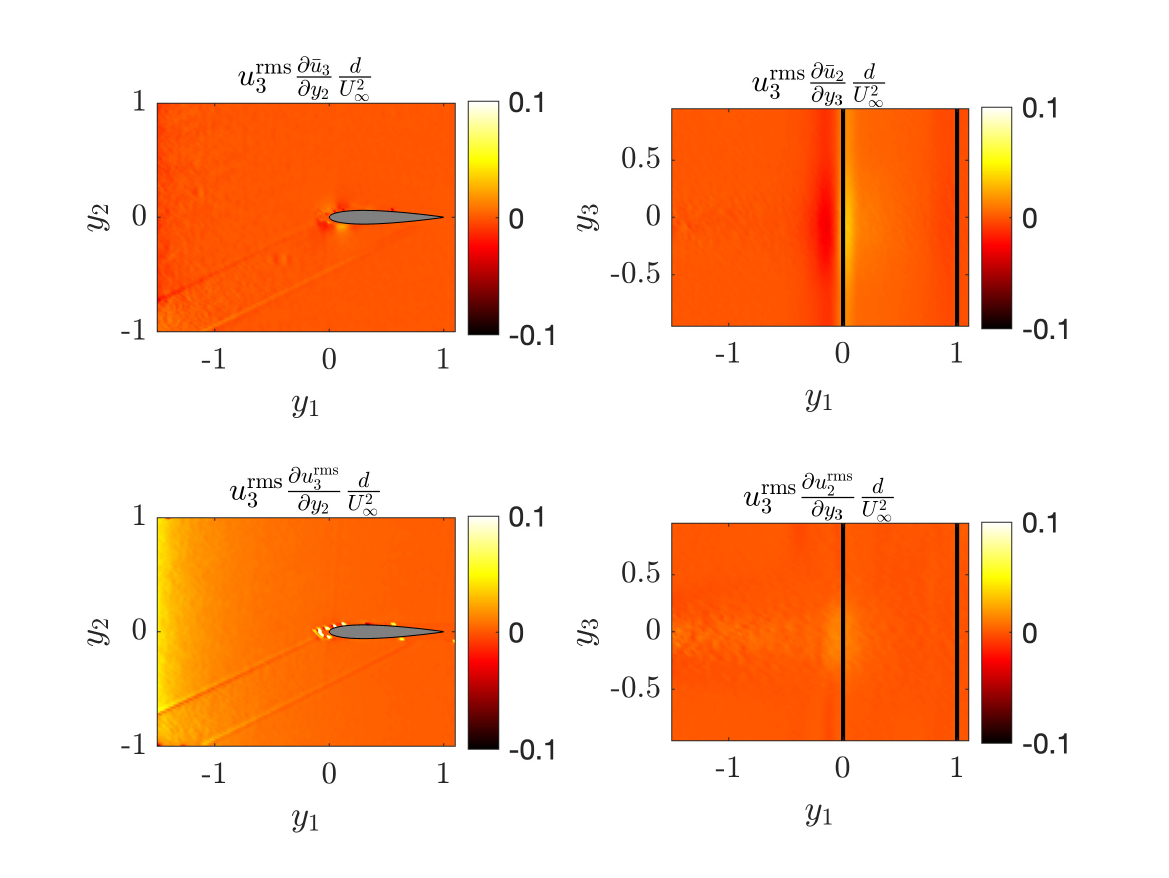}
    \caption{Estimation of the non-dimensional, unsteady Lamb-vector component ($\hat u_3 \omega_1 $) from data. The gradients of $u_2, u_3$ with respect to $y_3,y_2$ respectively are used to estimate the terms of \eqref{eq:streamwise_vort}. The left plots correspond to the $y_1-y_2$ plane where the gradient $\left( \frac{\partial }{\partial y_2}\right)$ of the $y_3-$ mean (top) and rms (bottom) velocity is computed. The right plots correspond to the $y_1-y_3$ plane where the gradient $\left( \frac{\partial }{\partial y_3}\right)$ of the $y_2-$ mean (top) and rms (bottom) velocity are computed. }
	\label{fig:Lamb}
\end{figure}

The current experimental set-up does not provide explicit information on the streamwise vorticity ($\omega_1$), however we can estimate its order of magnitude by considering the measurements taken in both planes. The streamwise vortcity is expressed as 
\begin{equation}
\omega_1 = \frac{\partial u_3}{\partial y_2}- \frac{\partial u_2}{\partial y_3},
\label{eq:streamwise_vort}
\end{equation}

\noindent while the Lamb-vector term associated with \eqref{eq:streamwise_vort} reads as,
\begin{equation}
\begin{aligned}
\omega_1 u_3 \frac{d}{U_{\infty}^2}\approx &\left(\frac{\partial \bar u_3}{\partial y_2}- \frac{\partial \bar u_2}{\partial y_3}\right) \frac{d}{U_{\infty}^2}\left( \bar u_3 + u^{\text{rms}}_3\right) \\
                      &+ \left(\frac{\partial u^{\text{rms}}_3}{\partial y_2}- \frac{\partial u^{\text{rms}}_2}{\partial y_3}\right)  \left( \bar u_3 + u^{\text{rms}}_3\right) \frac{d}{U_{\infty}^2}. \\                   
\end{aligned}
\label{eq:Lamb_omega1u3}
\end{equation}
\noindent Considering that $\bar u_3 \to 0$ near the airfoil it follows that 
\begin{equation}
\begin{aligned}
\omega_1 u_3 \frac{d}{U_{\infty}^2}\approx &\left(\frac{\partial \bar u_3}{\partial y_2}- \frac{\partial \bar u_2}{\partial y_3}\right) \frac{d}{U_{\infty}^2}u^{\text{rms}}_3 \\
                      &+ \left(\frac{\partial u^{\text{rms}}_3}{\partial y_2}- \frac{\partial u^{\text{rms}}_2}{\partial y_3}\right)   u^{\text{rms}} \frac{d}{U_{\infty}^2}. \\                   
\end{aligned}
\label{eq:Lamb_omega1u3v2}
\end{equation}
\noindent From figure \ref{meanV} it is observed that $\bar u_3$ is homogeneous (varies slowly) in $y_2$ and similarly $\bar u_2$ is homogeneous (varies slowly) in $y_3$. We thus expect their gradients $\left( \frac{\partial \bar u_3}{\partial y_2},\frac{\partial \bar u_2}{\partial y_3}\right) $ to become negligible and same reasoning can be followed for the airfoil-span-aligned, root-mean-square velocity $\left( \frac{\partial u^\mathrm{rms}_3}{\partial y_2} \to 0 \right)$. To demonstrate the latter, we compute the derivatives $\frac{\partial u_3}{\partial y_2}, \frac{\partial u_2}{\partial y_3}$ in the $y_1-y_2$ and $y_1-y_3$ planes respectively. Figure \ref{fig:Lamb} shows contour-plots of each term of  \eqref{eq:Lamb_omega1u3v2} and the corresponding orders of magnitude.

From the previous analysis we conclude that 
 \begin{equation}
\begin{aligned}
 \bar{\omega}_1 u^{\mathrm{rms}}_3 \frac{d^2}{U_\infty} &\approx o(\epsilon^2) \\
 \omega^{\mathrm{rms}}_1 u^{\mathrm{rms}}_3 \frac{d^2}{U_\infty}   &\approx  o(\epsilon^2)  \\
\frac{  \bar{\omega}_3 d }{U_{\infty}} \bar{u}_1 \frac{d^2}{U_\infty}   &\approx o(\epsilon)\\
 \omega^{\mathrm{rms}}_3  \bar{u}_1 \frac{d^2}{U_\infty}  &\approx o(1) \\
 \bar \omega_3 u^{\mathrm{rms}}_1 \frac{d^2}{U_\infty} &\approx o(\epsilon^2) \\
  \omega^{\mathrm{rms}}_3  u^{\mathrm{rms}}_1 \frac{d^2}{U_\infty}&\approx o(\epsilon) \\
\end{aligned}
\label{Scalings}
\end{equation}
 
 Neglecting the $\epsilon^2$-terms, eq. (\ref{App:Dipoles3}) reads as
 \begin{equation}
   \begin{aligned}
    p= -\rho \int_V   \left(\bar{\omega}_3 \bar{u}_1  + \hat \omega_3  \bar{u}_1+\bar{\omega}_3 \hat u_1 + \hat \omega_3 \hat u_1\right)   \frac{\partial G}{\partial y_2} \mathrm{d}V,
    \end{aligned}
     \label{App:Dipoles4}
\end{equation}

\begin{figure}
\centering
   \includegraphics[width=0.95\textwidth]{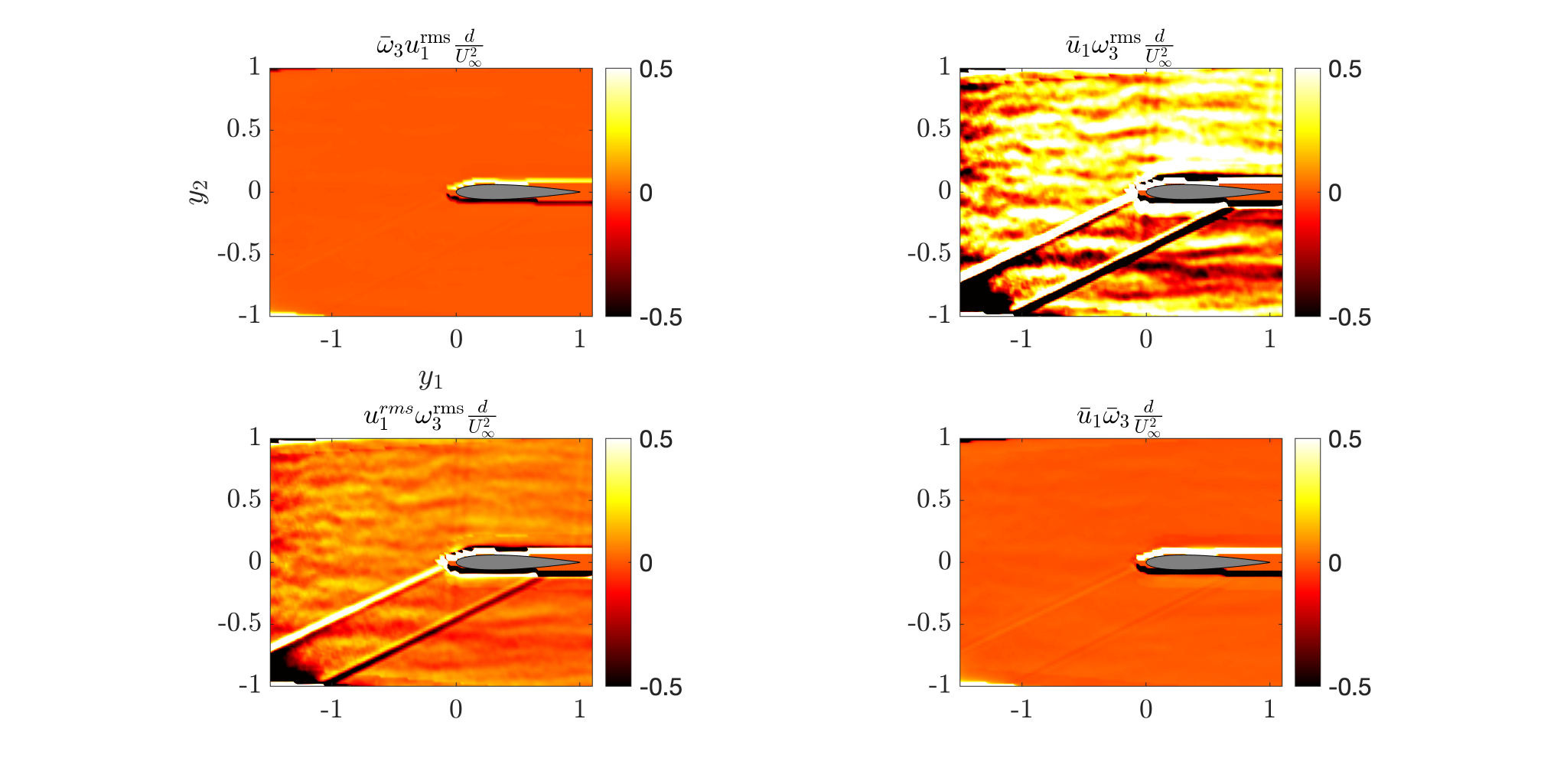}
    \caption{Comparison of the source-term components of Equation (\ref{App:Dipoles4}). The term $\bar{\omega}_3 \hat{u}_1$ is approximated by $\bar{\omega}_3 \hat{u}_1^{\mathrm{rms}}$ (top left), $\hat{\omega}_3 \hat{u}_1$ by $\omega_3^{\mathrm{rms}} u_1^{\mathrm{rms}}$ (bottom left), and $\hat{\omega}_3 \bar{u}_1$ by $\omega_3^{\mathrm{rms}} \bar{u}_1$ (top right), while $\bar{\omega}_3 \bar{u}_1$ is shown in the bottom-right plot. All quantities are non-dimensionalised by $\frac{d^2}{U_{\infty}}$.}
	\label{fig:Lamb_omega3}
\end{figure}

The remaining terms of \eqref{App:Dipoles4} are compared on figure \ref{fig:Lamb_omega3}. Notice that the measurements at the region near the boundary layer is excluded as discussed in section \ref{sec:Experiments} and the source term can be approximated to first order by the term $\hat \omega_3  \bar{u}_1$.  It thus follows,  
 \begin{equation}
   \begin{aligned}
    p=- \rho \int_V \hat  \omega_3  \bar{u}_1  \frac{\partial G}{\partial y_2} \mathrm{d}V.
    \end{aligned}
     \label{App:Dipoles5}
\end{equation}

\newpage
\section{Validation of the source-panel-method algorithm}
\label{panel}

In this appendix we validate the source-panel-method algorithm used to compute
the Kirchhoff-vector components appearing in \eqref{pot_flow}, and thereby
obtain the spatial derivatives of the compact Green function \eqref{newGF}. The Kirchhoff vector and the spatial derivatives of its components
(i.e.\ the velocity potential and velocity field) are computed for a circular cylinder of
radius $R=1/2$ (non-dimensional units). These numerical results are compared with the classical analytical
solution \citep[p.~60]{Howe2002TheoryOV}:
\begin{equation}
    Y_{1,2}^{\mathrm{an}}
    = y_{1,2}\left( 1 + \frac{R}{y_1^2 + y_2^2} \right).
    \label{cyli_kir}
\end{equation}

\begin{figure}
    \centering
    \includegraphics[width=0.6\textwidth]{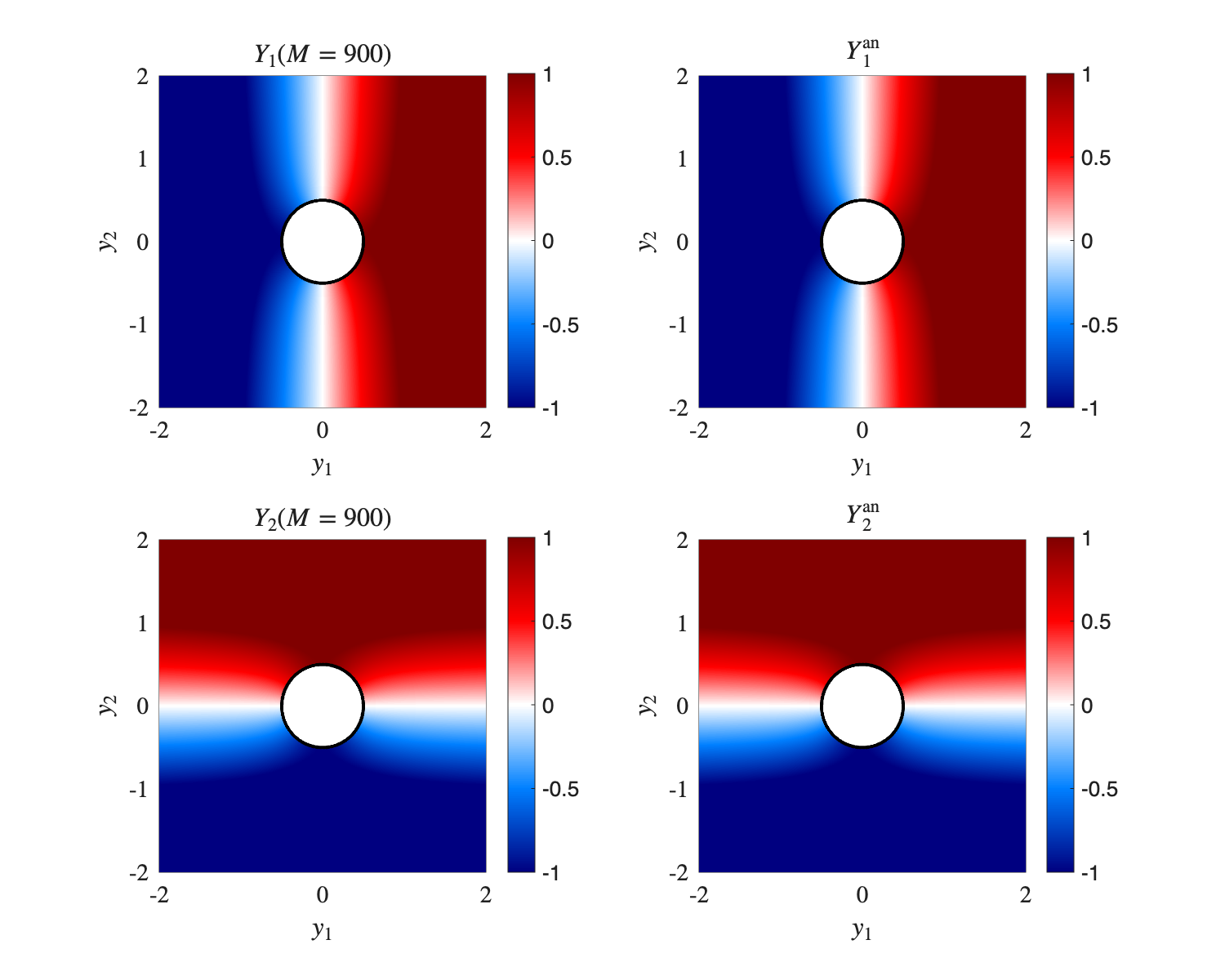}
    \caption{Computed Kirchhoff vector components using 900 panels (left)
    compared with the analytical expression of \eqref{cyli_kir}.}
    \label{fig:Kir}
\end{figure}

\begin{figure}
    \centering
    \includegraphics[width=0.6\textwidth]{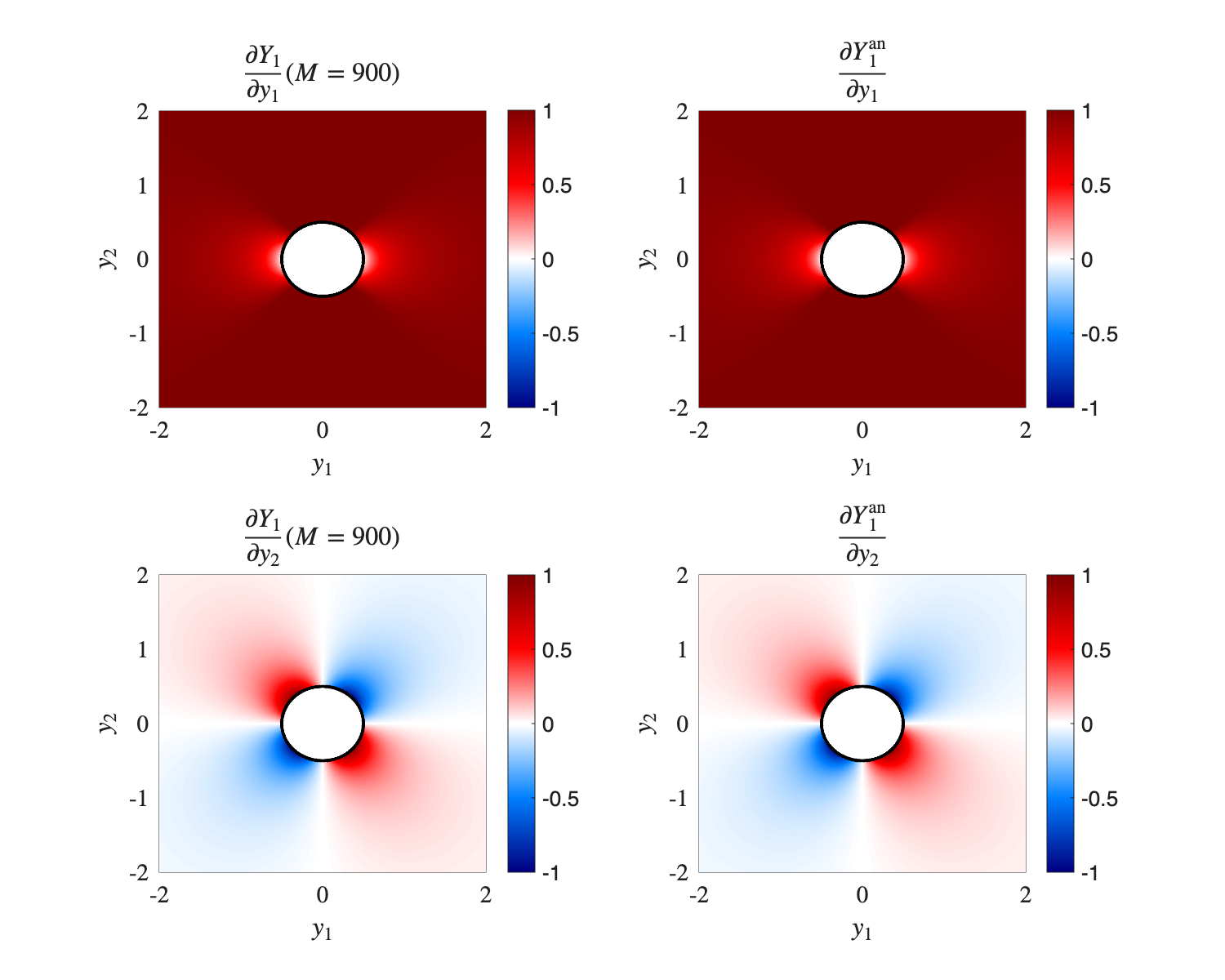}
    \caption{Computed derivatives of the Kirchhoff vector component $Y_1$
    using 900 panels (left) compared with the analytical $y_1$-derivative
    of \eqref{cyli_kir} (right).}
    \label{fig:dY1}
\end{figure}

\begin{figure}
    \centering
    \includegraphics[width=0.6\textwidth]{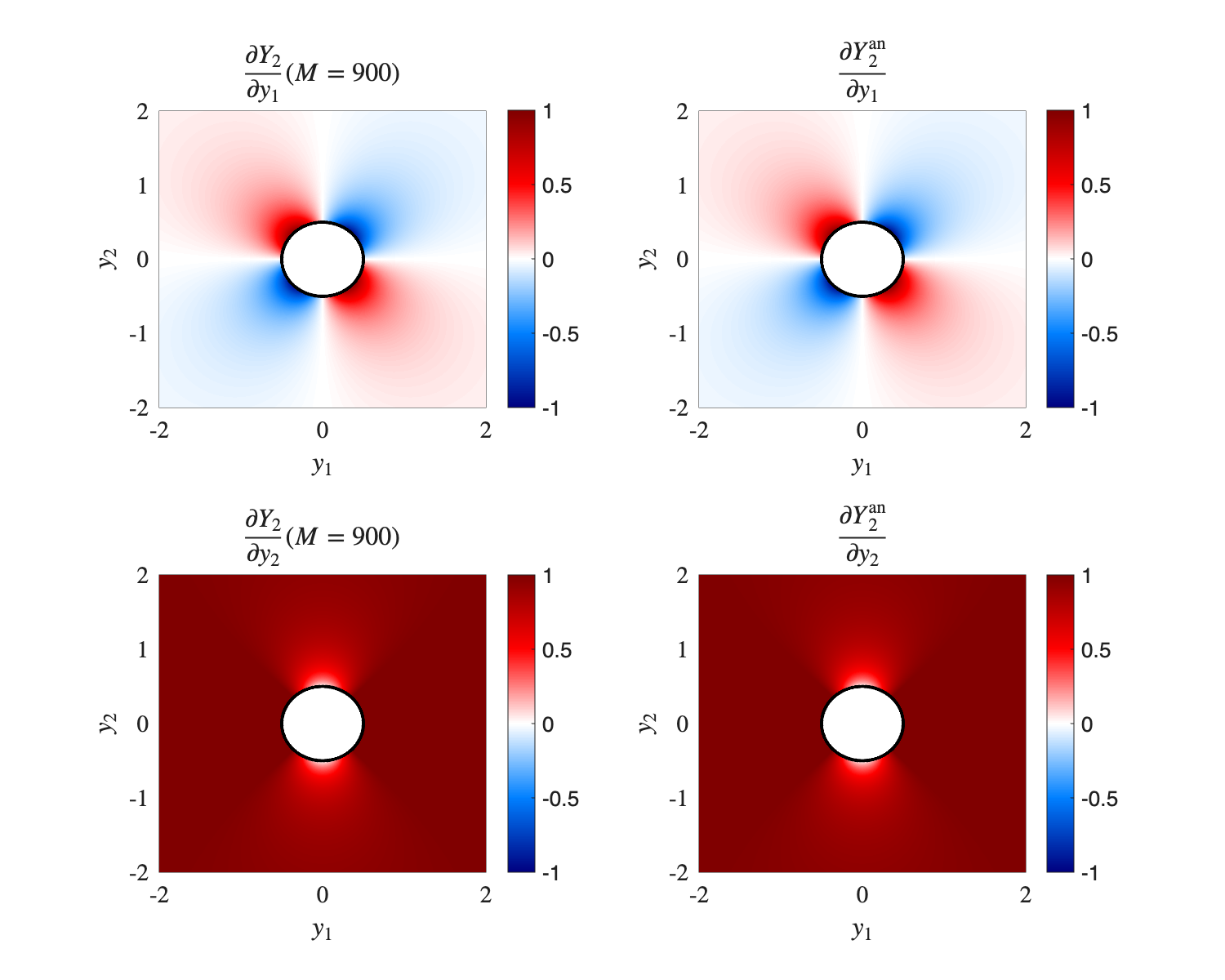}
    \caption{Computed derivatives of the Kirchhoff vector component $Y_2$
    using 900 panels (left) compared with the analytical $y_2$-derivative
    of \eqref{cyli_kir} (right).}
    \label{fig:dY2}
\end{figure}

Figures~\ref{fig:Kir}--\ref{fig:dY2} show the computed Kirchhoff vector
components and their spatial derivatives obtained using $M=900$ panels,
together with the corresponding analytical expressions obtained from
\eqref{cyli_kir}. The good agreement between the numerical and analytical Kirchhoff vectors validates the
source-panel-method algorithm.

\newpage
\section{Influence of window function}
\label{App:Convergence}

\begin{figure}
\centering
   \includegraphics[width=0.95\textwidth]{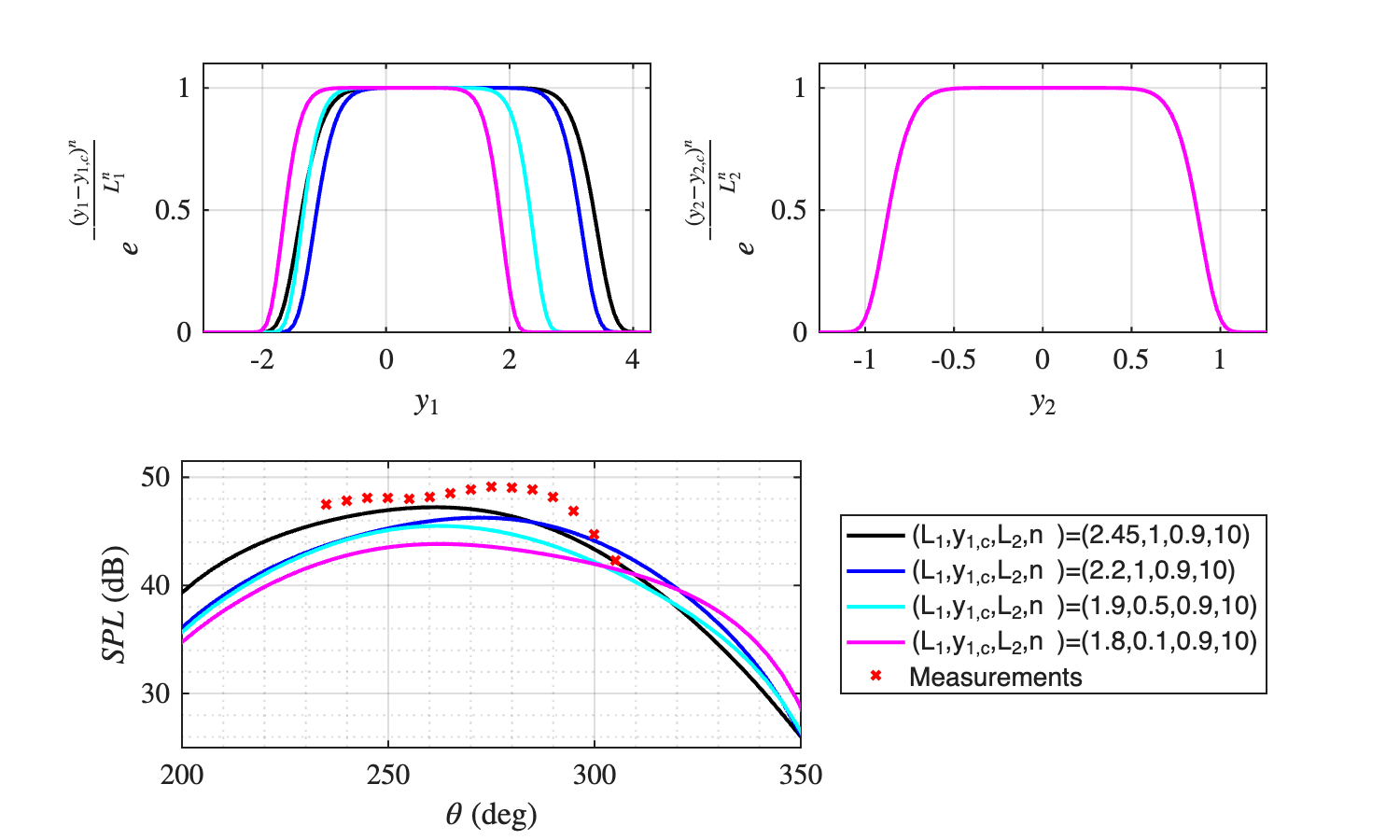}
    \caption{Influence of the window function of eq \eqref{eq:super_Gaussian} on the acoustic field.}
\label{fig:window}
\end{figure}

To asses the influence of the window function we compute the acoustic field by varying the parameters of eq \eqref{eq:super_Gaussian}. The results are presented in figure \ref{fig:window} demonstrating that the larger windows ameliorate the agreement between the estimated and measured acoustic spectra. 

\newpage
\section{Convergence study of wavenumber transform along $y_2$}
\label{App:k2_spectra}
\begin{figure}
\centering
   \includegraphics[width=0.95\textwidth]{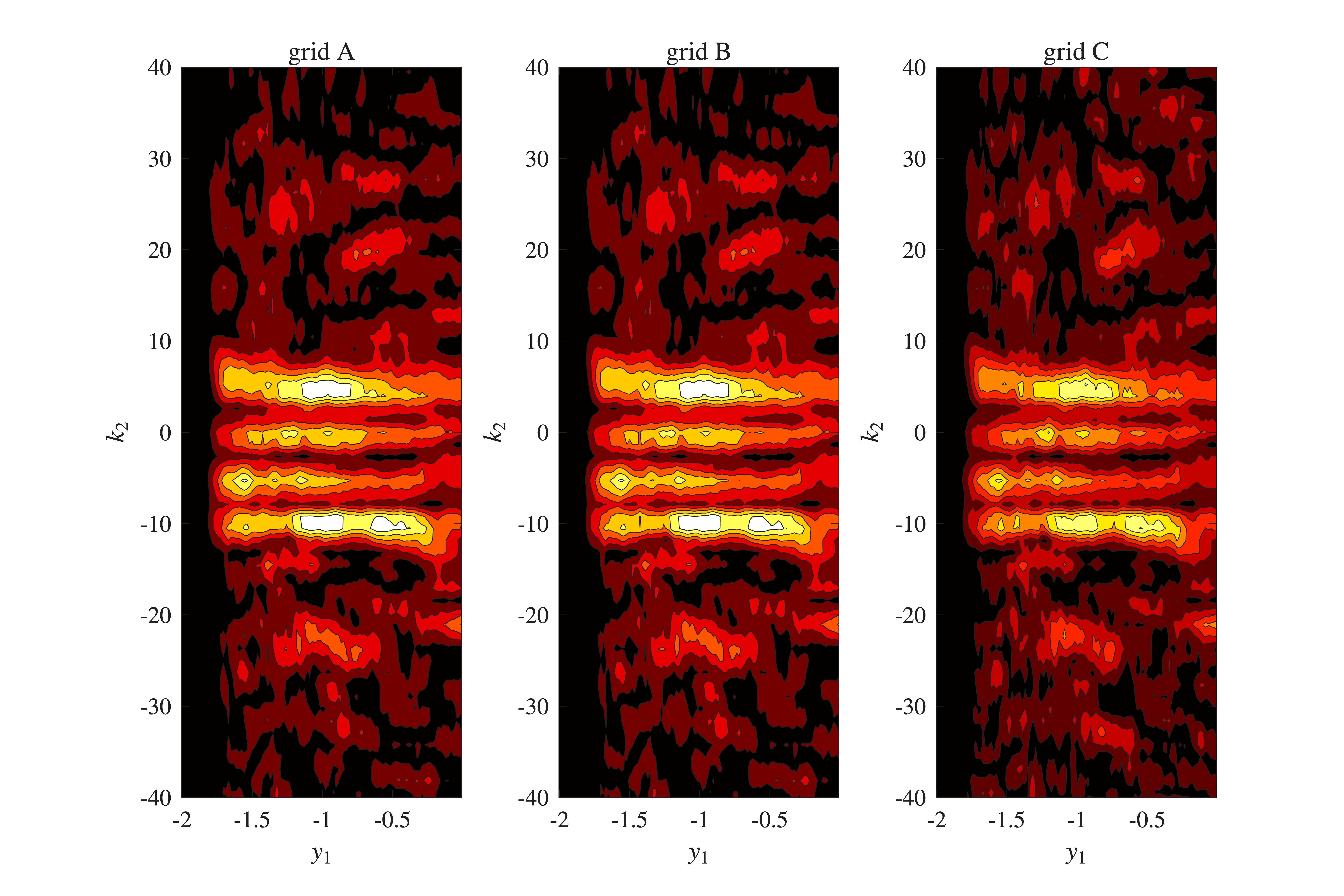}
    \caption{Amplitude of the wavenumber-transformed source term in $y_2$ $\left(  \left|   \hat  {\hat{S}}\left( y_1, k_2 \right) \right| \right) $ for the grids A,B,C shown in table \ref{tab:k2}.}
\label{fig:k2_spectra_conv}
\end{figure}

 To ensure convergence of the $k_2-$spectra we compute the wavenumber transform on two decimated grids and compare to the results of \S \ref{subsec:wavenumber_transform}. Table \ref{tab:k2} shows the number of  grid points in $y_2$ ($N_{y_2})$, the number of Fourier-transform points ($N_{y_2,\mathrm{fft}}$) used after zero-padding the signal, the spatial discretisation $\left( \Delta y_2 \right)$ and the minimum wavelength that is resolved $\left( \lambda_{\mathrm{min}} \right)$. The amplitude of the wavenumber-transformed source is shown in figure \ref{fig:k2_spectra_conv} for the three grids of table \ref{tab:k2}. We can therefore conclude that the $k_2$- spectrum converges. 
 
 \begin{table}
  \centering
  \begin{tabular}{ccccc}
    \hline
    Case & $N_{y_2}$& $N_{y_2,\mathrm{fft}}$ &  $\Delta y_2$ &  $\lambda_{\mathrm{min}}$ \\
    \hline
    Grid A & 135 & 256 &0.019 &$0.19d$ \\
    Grid B & 68 & 128 & 0.037& $0.37 d$ \\
        Grid C & 34 & 64 & 0.07 & $0.7 d$\\
    \hline
  \end{tabular}
    \caption{Grid parameters used for the decimated $y_2$–grids and the corresponding wavenumber discretisation $\Delta k_2$ and sampling frequency $\lambda_{2_S}$.}
  \label{tab:k2}
\end{table}

\newpage
\bibliographystyle{jfm}
\bibliography{refs.bib}

\end{document}